\documentclass[aps,prl,reprint,superscriptaddress]{revtex4-2}
\usepackage{graphicx}
\usepackage{amsmath,bm}
\usepackage{amssymb}
\usepackage{commath} 
\usepackage{xcolor}

\begin{document}

\title{Conformation-Mediated Kinetics of Polymer Chain Scission under Tension}
\author{Jie Zhu}
\affiliation{Department of Engineering Science, University of Oxford, Oxford OX1 3PJ, United Kingdom.}

\author{Laurence Brassart}
\affiliation{Department of Engineering Science, University of Oxford, Oxford OX1 3PJ, United Kingdom.}

\begin{abstract}
Chain scission is a key molecular process underlying damage and fracture in polymer networks. In this Letter, we develop a statistical-mechanical framework for predicting chain-scission kinetics while accounting for three-dimensional (3D) conformational fluctuations. Within transition-state theory, scission is formulated as a multichannel first-rupture problem, with bond-specific rates governed primarily by self-consistent potentials of mean force. In the freely jointed limit, the additional 3D configurational freedom enhances rupture relative to the collinear 1D reference. Finite bending stiffness introduces orientational correlations that can reverse this enhancement and, at high stiffness, reduce rupture rates by orders of magnitude. These correlations also make rupture bond-position dependent, with higher rates near the chain ends and a common interior rate. For sufficiently long chains, the interior contribution dominates, yielding linear scaling of the chain-scission rate with chain length. These molecularly resolved rates provide physically grounded inputs for future network-scale models of polymer damage and fracture.
\end{abstract}

\maketitle

\begin{figure}
    \includegraphics{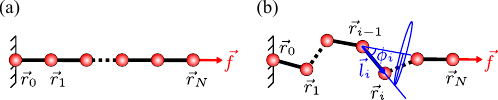}
    \caption{Schematic of a polymer chain of $N+1$ atoms, with one end fixed at $\vec{r}_0 = \vec{0}$ and a constant force $\vec{f}$ applied to the free end at $\vec{r}_N$. (a) 1D bead-string approximation, in which all backbone bonds are collinear with the applied force and identically loaded. (b) 3D chain model with deformable bond lengths and bond angles.}
    \label{fig:schematicofchain}
\end{figure}

Chain scission, the irreversible rupture of covalent backbone bonds, is a key molecular process underlying damage and fracture in polymer networks \cite{beyer2005mechanochemistry,ducrot2014toughening,slootman2020quantifying,wang2023investigating}. Under tension, rupture of a single bond is typically described within transition-state theory (TST) as a thermally activated barrier-crossing process in which the applied force lowers the activation barrier and thereby increases the bond rupture rate \cite{kramers1940brownian,dudko2006intrinsic,freund2009characterizing,dudko2009single,friddle2012interpreting}. Extending this single-bond picture to a polymer chain is nontrivial because the applied force is transmitted through many backbone bonds whose orientations and extensions fluctuate collectively in three dimensions (3D). Traditional one-dimensional (1D) chain-scission models simplify this problem by idealizing the backbone as a sequence of collinear, identically loaded bonds (Fig. \ref{fig:schematicofchain}(a)) \cite{sebastian1999breaking,puthur2002theory,charan2021aging,theodorou2024simple}. To include 3D conformational effects, related approaches introduce an effective force acting on a representative bond and use it in a single-bond TST expression to obtain the bond rupture rate $k_\mathrm{bond}$ \cite{lavoie2019modeling,guo2021micromechanics}. In both 1D models and such mean-field 3D descriptions, chain scission is approximated as the first rupture event among $N$ statistically equivalent failure sites, leading to the chain scission rate $k_\mathrm{chain} \approx N k_\mathrm{bond}$. While mathematically convenient, these treatments do not resolve how 3D conformational fluctuations reshape the local free-energy landscape of individual bonds and thereby modify their rupture kinetics.

To address this limitation, we develop a bond-resolved statistical-mechanical TST framework for chain scission in a 3D polymer chain with deformable bond lengths and bond angles under constant tension (Fig. \ref{fig:schematicofchain}(b)). We treat rupture of each backbone bond as a distinct channel, giving the total rate as $k_\mathrm{chain} = \sum_i k_{\mathrm{bond},i}$. Each bond-resolved rate is governed primarily by its potential of mean force (PMF), which describes the effective free-energy landscape for stretching that bond while all other bonds remain intact. We evaluate the PMFs using a transfer-matrix (TM) scheme that captures nearest-neighbor orientational correlations and determine the rupture thresholds self-consistently from their barrier tops. In the freely jointed (FJ) limit, the additional 3D configurational freedom yields higher rupture rates than the collinear 1D reference, with this enhancement diminishing as the force increases. Finite bending stiffness introduces orientational correlations that can reverse this enhancement and, at high stiffness, reduce rupture rates by orders of magnitude below the 1D value. These correlations also make rupture bond-position dependent: bonds near the chain ends rupture more readily, while interior bonds approach a common rate $k_\infty$. For sufficiently long chains, the interior contribution dominates, giving $k_\mathrm{chain} \sim N k_\infty$. We next specify the chain model and develop the corresponding TST formulation.

We consider a polymer chain of $N+1$ atoms at positions $\vec r_i$ $(i=0,\ldots,N)$, connected by $N$ backbone bonds with bond vectors $\vec{l}_i = \vec{r}_i-\vec{r}_{i-1}$. Atom $0$ is fixed at $\vec{r}_0 = \vec{0}$, and a constant force $\vec{f}$ is applied to atom $N$ (Fig. \ref{fig:schematicofchain}(b)). The bond length is $l_i = \|\vec{l}_i\|$, and the polar angle of $\vec l_i$ with respect to the direction of $\vec f$ is denoted by $\theta_i$. To describe bond stretching and rupture, we use a Morse potential $v_\mathrm{str}(l)=D_e[1 - e^{-a(l-l_e)}]^2$, where $D_e$ is the dissociation energy, $a$ is the inverse interaction range, and $l_e$ is the equilibrium bond length. Bond-angle deformations are penalized by $v_\mathrm{ben}(\phi)=k_\phi(\phi - \phi_e)^2/2$, where $\phi_i = \angle(\vec l_{i-1},\vec l_i)$ is the local bond angle, $k_\phi$ is the bending stiffness, and $\phi_e$ is the equilibrium bond angle. Torsional interactions are neglected, corresponding to freely rotating dihedrals. Nonbonded intrachain interactions are also neglected. The configurational energy is
\begin{equation}
    U = \sum_{i=1}^{N}v_\mathrm{str}{(l_i)} + \sum_{i=2}^{N}v_\mathrm{ben}{(\phi_i)} - \sum_{i=1}^N fl_i \cos{\theta_i},
\end{equation}
where $f = \|\vec{f}\|$. The force term follows from $-\vec{f} \cdot \vec r_N$, with $\vec r_N=\sum_{i=1}^N\vec l_i$. Details of the model setup are given in the Supplemental Material, Sec. S1.1 \cite{supplemental}.

Chain scission is formulated as a multichannel first-rupture problem in bond-length space. Each bond $i$ is assigned a rupture threshold $l_i^\ddagger$ and is regarded as ruptured when $l_i \geq l_i^\ddagger$, with the threshold set denoted by $\boldsymbol{l}^\ddagger = \{l_i^\ddagger\}_{i=1}^N$. The chain remains intact only when all bonds satisfy $l_i < l_i^\ddagger$, whereas crossing any bond threshold places the configuration in the dissociated region. The boundary between these regions defines the dividing surface $\Sigma(\boldsymbol{l}^\ddagger)$. Within TST, the chain-scission rate is estimated as the positive equilibrium flux through $\Sigma$ from the intact basin to the dissociated region, normalized by the equilibrium population of the intact basin \cite{zwanzig2001nonequilibrium}. We identify $N$ bond-resolved rupture channels
\begin{equation}
    \Sigma_i(\boldsymbol{l}^\ddagger) = \{l_i = l_i^\ddagger,~ l_j < l_j^\ddagger\,\text{for all}\,j \neq i\}.
\end{equation}
Each $\Sigma_i$ contains configurations in which bond $i$ is the unique bond at its rupture threshold while all other bonds remain intact. Simultaneous threshold crossings of two or more bonds make no contribution to the TST flux (Supplemental Material, Sec. S1.2 \cite{supplemental}), so the total positive flux decomposes additively over the rupture channels. Dividing each channel flux by the common intact-basin population defines the bond-resolved rate $k_{\mathrm{bond},i}(\boldsymbol{l}^\ddagger;f)$, and the chain-scission rate becomes
\begin{equation}
    k_{\mathrm{chain}}(\boldsymbol{l}^\ddagger;f) = \sum_{i=1}^N k_{\mathrm{bond},i}(\boldsymbol{l}^\ddagger;f).
\end{equation}

To evaluate $k_{\mathrm{bond},i}$, we integrate out the Maxwell-Boltzmann momenta and obtain a form that factorizes into a kinetic prefactor and a configurational contribution (Supplemental Material, Secs. S1.3--S1.4 \cite{supplemental}),
\begin{equation}
\label{eq:k_bond_with_W}
    k_{\mathrm{bond},i}(\boldsymbol{l}^\ddagger; f) = \frac{1}{\sqrt{2\pi m_i^l \beta}} \frac{\displaystyle e^{-\beta \mathcal{W}_i(l_i^\ddagger; \boldsymbol{l}_{-i}^\ddagger, f)}}{\displaystyle \int_0^{l_i^\ddagger} e^{-\beta \mathcal{W}_i(l; \boldsymbol{l}_{-i}^\ddagger, f)}\,dl},
\end{equation}
where $\beta \equiv (k_BT)^{-1}$, with $k_B$ the Boltzmann constant and $T$ the temperature. The prefactor $1/\sqrt{2\pi m_i^l\beta}$ is the thermal average of the positive normal velocity through $\Sigma_i$, and $m_i^l$ is the effective mass associated with the bond-length coordinate $l_i$. For identical atomic masses $m$, the fixed anchor makes the first bond special, giving $m_1^l = m$, whereas $m_i^l = m/2$ for $i \geq 2$. Thus, all bonds except the first share the same kinetic prefactor. 

The configurational contribution is governed by a bond-length PMF $\mathcal{W}_i(l; \boldsymbol{l}_{-i}^\ddagger, f)$, where $\boldsymbol{l}_{-i}^\ddagger = \{l_j^\ddagger\}_{j \neq i}$ denotes the thresholds of all bonds except $i$. We define $\mathcal{W}_i(l; \boldsymbol{l}_{-i}^\ddagger, f) = -{\beta}^{-1} \ln{\mathcal{G}_i(l; \boldsymbol{l}_{-i}^\ddagger, f)}$, where $\mathcal{G}_i(l; \boldsymbol{l}_{-i}^\ddagger, f)$ is the constrained configurational weight at fixed $l_i = l$, with all other bonds satisfying $l_j<l_j^\ddagger$ for $j \neq i$. This PMF is the effective free-energy landscape for stretching bond $i$ while the remaining chain stays intact, and its use in TST follows the standard PMF-based representation \cite{watney2006calculation,schenter2003generalized}. Accordingly, the configurational factor in Eq. \eqref{eq:k_bond_with_W} is the normalized Boltzmann probability density associated with $\mathcal W_i$, evaluated at $l_i^\ddagger$ after normalization over $0 < l < l_i^\ddagger$. Any $l$-independent additive constant in $\mathcal W_i$ therefore cancels out. Apart from the distinct kinetic prefactor of bond $1$, bond-to-bond rate variations are thus encoded entirely in $\mathcal W_i$.

Because the bending interaction couples neighboring bond orientations, the force dependence of $\mathcal{W}_i(l; \boldsymbol{l}_{-i}^\ddagger, f)$ is modulated by the local conformational environment of the selected bond. Building on our previous TM formulations for chain elasticity \cite{zhu2025stretching,zhu2026elasticity}, we compute this PMF by propagating orientational correlations along the chain while keeping the selected bond length $l_i = l$ explicit (Supplemental Material, Sec. S2.1 \cite{supplemental}). Dropping $l$-independent constants, the resulting PMF is
\begin{equation}
\label{eq:PMF_with_angular_weight}
    \mathcal{W}_i(l) = v_\mathrm{str}(l) - {\beta}^{-1} \ln{\left[l^2 \int_0^\pi w_i(\theta) e^{\beta fl\cos{\theta}} \sin{\theta}\,d\theta \right]},
\end{equation}
where, for compactness, we write $\mathcal{W}_i(l)$ and $w_i(\theta)$ for $\mathcal{W}_i(l; \boldsymbol{l}_{-i}^\ddagger, f)$ and $w_i(\theta; \boldsymbol{l}_{-i}^\ddagger, f)$, respectively. The angular weight $w_i(\theta)$, constructed from forward and backward TM messages (Supplemental Material, Sec. S2.2 \cite{supplemental}), encodes the contribution of the surrounding chain to the statistical weight when bond $i$ has polar angle $\theta$. The remaining factors have clear physical interpretations: $v_\mathrm{str}(l)$ is the bare stretching energy, $l^2\sin\theta$ is the Jacobian for the 3D bond-vector measure in spherical coordinates, and $e^{\beta f l\cos\theta}$ accounts for the direct coupling of bond $i$ to the applied force. 

In the FJ limit, $v_\mathrm{ben}(\phi) = 0$, so bond orientations are independent and the surrounding-chain dependence disappears, giving $w_\mathrm{FJ}(\theta) = 1$. Eq. (\ref{eq:PMF_with_angular_weight}) then reduces to (Supplemental Material, Sec. S3.1 \cite{supplemental})
\begin{equation}
    \mathcal W_\mathrm{FJ}(l;f) = v_\mathrm{str}(l) - {\beta}^{-1} \ln{\left[2l^2\frac{\sinh{(\beta fl)}}{\beta fl} \right]}.
\end{equation}
For comparison, in a collinear 1D reference, each bond is constrained along the force direction, giving $\mathcal W_\mathrm{1D}(l;f) = v_\mathrm{str}(l) - fl$ (Supplemental Material, Sec. S3.2 \cite{supplemental}).

For the default numerical calculations, we use the carbon--carbon-based parameters of Refs. \cite{shannon2019anharmonic,sorensen1988prediction} at $T = 293.15~\mathrm{K}$: $al_e=2.15$, $\beta D_e=279$, $\phi_e = 69^\circ$, and $\beta k_\phi\pi^2 = 1.82 \times 10^3$ (Supplemental Material, Sec. S4 \cite{supplemental}). The dimensionless force $f l_e/D_e$ and chain length $N$ are varied as indicated; stiffness-dependent comparisons additionally include $\beta k_\phi\pi^2 = 10^2$ and $10^4$.

\begin{figure}
    \includegraphics{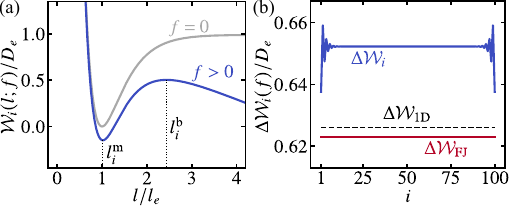}
    \caption{Bond-length PMFs and bond-resolved activation barriers. (a) Representative zero-force PMF and self-consistent finite-force PMF at $fl_e/D_e = 0.2$ for an interior bond ($i=50$). (b) Bond-resolved activation barriers $\Delta \mathcal{W}_i(f)$ at the same force, compared with the FJ limit and the 1D reference. The finite-bending chain has $N=100$ and $\beta k_\phi\pi^2 = 1.82 \times 10^3$.}
    \label{fig:representative_PMF}
\end{figure}

We first examine representative bond-length PMFs in Fig. \ref{fig:representative_PMF}(a). At zero force, global rotational symmetry makes the angular weight independent of $\theta$, $\boldsymbol{l}_{-i}^\ddagger$, and the bond index. Its constant value contributes only an $l$-independent shift to the PMF; we therefore take $w_i(\theta;0)=1$ for every $i$. Consequently, all bonds share the same zero-force PMF, $\mathcal{W}(l;0) = v_\mathrm{str}(l)-\beta^{-1}\ln (2 l^2)$. Although the logarithmic entropic term gives rise to a formal barrier at large $l$, for $\beta D_e \gg 1$ this barrier lies far beyond the range shown, so the zero-force profile appears effectively single-welled. At finite force, the constrained PMFs become bond specific, each exhibiting a bonded minimum and a barrier top along the bond-length coordinate. The minimum identifies the most probable length of bond $i$ within the intact basin. Following the variational-TST principle of optimizing the dividing surface \cite{truhlar1980variational,makarov2015single}, we take the barrier top as the PMF-based rupture threshold. Because $\mathcal{W}_i(l; \boldsymbol{l}_{-i}^\ddagger, f)$ depends parametrically on the thresholds of the other bonds, these thresholds must be determined self-consistently by iterating between PMF evaluations and barrier-top updates until convergence (Supplemental Material, Sec. S2 \cite{supplemental}). After convergence, we denote the self-consistent PMF by $\mathcal{W}_i(l; f)$ and its bonded-minimum and barrier-top locations by $l_i^\mathrm{m}(f)$ and $l_i^\mathrm{b}(f)$, respectively. Substituting $\boldsymbol{l}_\mathrm{opt}^\ddagger(f) = \{l_i^\mathrm{b}(f)\}_{i=1}^N$ and the PMFs $\mathcal{W}_i(l; f)$ into Eq. (\ref{eq:k_bond_with_W}) gives the optimal bond-resolved rates $k_{\mathrm{bond},i}^\mathrm{opt}(f)$, whose sum yields the optimal chain-scission rate $k_\mathrm{chain}^\mathrm{opt}(f) = \sum_{i=1}^N k_{\mathrm{bond},i}^\mathrm{opt}(f)$. This PMF-based TST analysis is restricted to $f < f_c$, where $f_c$ is the chain-level critical force at which the bonded minimum and barrier top first coincide for at least one bond (Sec. S1.4).

To interpret these rates in the familiar Arrhenius picture, we define the activation barrier for bond $i$ as $\Delta \mathcal{W}_i(f) = \mathcal{W}_i(l_i^\mathrm{b}(f);f) - \mathcal{W}_i(l_i^\mathrm{m}(f);f)$. In the high-barrier regime, $\beta \Delta \mathcal{W}_i(f) \gg 1$, applying a harmonic approximation to the PMF about the bonded minimum reduces the full TST expression in Eq. (\ref{eq:k_bond_with_W}) to
\begin{equation}
    \label{eq:Arrhenius_form}
    k_{\mathrm{bond},i}^\mathrm{opt}(f) \approx \frac{\omega_i^\mathrm{m}(f)}{2\pi} e^{-\beta \Delta \mathcal{W}_i(f)}, 
\end{equation}
where $\omega_i^\mathrm{m}(f)$ is defined by the local PMF curvature, $m_i^l[\omega_i^\mathrm{m}(f)]^2 = \left.{\partial^2 \mathcal{W}_i / \partial l^2}\right\vert_{l = l_i^\mathrm{m}(f)}$ \cite{hanggi1990reaction}. The prefactor ${\omega_i^\mathrm{m}(f)}/{(2\pi)}$ sets the characteristic attempt frequency, whereas the Arrhenius factor captures the dominant exponential dependence on $\Delta \mathcal{W}_i(f)$. 

We next compare the bond-resolved activation barriers $\Delta \mathcal{W}_i(f)$ at $fl_e/D_e = 0.2$ with the bond-independent FJ limit and 1D reference in Fig. \ref{fig:representative_PMF}(b). The FJ barrier lies below the 1D value, $\Delta \mathcal W_\mathrm{FJ}(f) < \Delta \mathcal W_\mathrm{1D}(f)$, reflecting the additional configurational entropy arising from the 3D bond-vector measure and orientational sampling. At finite bending stiffness, neighboring bond orientations become correlated. Truncation of these correlations at the chain ends makes the angular weights $w_i(\theta)$ position dependent, while the homogeneous local interactions and identical truncation at both ends ensure mirror symmetry about the chain midpoint, $w_i(\theta) = w_{N+1-i}(\theta)$ (Supplemental Material, Sec. S5.1 \cite{supplemental}). Through Eq. (\ref{eq:PMF_with_angular_weight}), this bond-position dependence is inherited by the activation barriers, which consequently satisfy $\Delta \mathcal W_i(f) = \Delta \mathcal W_{N+1-i}(f)$. Each terminal bond receives orientational constraints from only one adjacent subchain, whereas the first few nonterminal bonds are coupled to two subchains of unequal lengths that impose different orientational constraints. As the influence of the chain ends decays toward the interior, the activation barriers approach the interior plateau $\Delta \mathcal W_\infty(f)$ through damped oscillations, with their lowest values occurring at the terminal bonds. Relative to the FJ limit, bending correlations suppress independent bond alignment with the force and favor bending-compatible orientations, raising $\Delta \mathcal W_i(f)$ above $\Delta \mathcal W_\mathrm{FJ}(f)$ throughout the chain. For $\beta k_\phi\pi^2 = 1.82 \times 10^3$, the entire barrier profile lies above even $\Delta \mathcal W_\mathrm{1D}(f)$, showing that bending constraints can more than offset the entropic barrier lowering in the FJ limit.

\begin{figure}
    \includegraphics{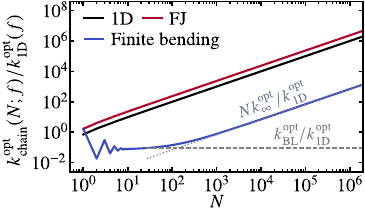}
    \caption{Finite-size scaling of chain-scission kinetics. Chain-scission rate $k_{\mathrm{chain}}^\mathrm{opt}(N;f)$, normalized by the common 1D bond rate $k_{\mathrm{1D}}^\mathrm{opt}(f)$, versus chain length $N$ at $fl_e/D_e = 0.2$, comparing the 1D reference, FJ limit, and finite-bending case with $\beta k_\phi\pi^2 = 1.82 \times 10^3$. For the finite-bending result, the horizontal dashed line marks the total contribution from the two converged boundary layers, $k_\mathrm{BL}^\mathrm{opt}(f) / k_{\mathrm{1D}}^\mathrm{opt}(f)$, while the dotted line shows the leading bulk asymptote, $N k_\infty^\mathrm{opt}(f) / k_{\mathrm{1D}}^\mathrm{opt}(f)$.}
    \label{fig:chain_length_dependence}
\end{figure}

These bond-resolved barrier trends are reflected in the corresponding rates and, consequently, in the finite-size scaling of the chain-scission rate. In the 1D reference and FJ limit, all bonds $i \geq 2$ share the common rates $k_\mathrm{1D}^\mathrm{opt}(f)$ and $k_\mathrm{FJ}^\mathrm{opt}(f)$, respectively, while bond $1$ is slower by a factor of $1/\sqrt{2}$ because of its distinct kinetic prefactor. Fig. \ref{fig:chain_length_dependence} shows the normalized chain-scission rate $k_{\mathrm{chain}}^\mathrm{opt}(N;f) / k_{\mathrm{1D}}^\mathrm{opt}(f)$ versus chain length $N$ at $fl_e/D_e = 0.2$ for the 1D reference, FJ limit and finite-bending case with $\beta k_\phi\pi^2 = 1.82 \times 10^3$. The 1D and FJ chain rates grow linearly with $N$, and $k_\mathrm{FJ}^\mathrm{opt}(f) > k_\mathrm{1D}^\mathrm{opt}(f)$, consistent with the lower FJ activation barrier discussed above. For the finite-bending case, the result at $N = 1$ coincides with the FJ limit because no bond angle exists and the bending energy is absent. Bending correlations first enter at $N = 2$, and for $N \geq 3$ their truncation at the chain ends generates a nonuniform barrier profile (Supplemental Material, Sec. S6.1 \cite{supplemental}). As $N$ increases, the near-end barrier profiles approach their long-chain forms, while a bulk-like interior plateau $\Delta \mathcal W_\infty(f)$ develops. The spatial extent of the converged end-induced perturbation is characterized by the boundary-layer length $\ell_b(f)$. We denote the common interior bond rate by $k_\infty^\mathrm{opt}(f)$ and the summed rate contribution from the two converged boundary layers by $k_\mathrm{BL}^\mathrm{opt}(f)$. For $N > 2 \ell_b(f)$, the chain rate is approximately $k_\mathrm{chain}^\mathrm{opt}(N; f) \simeq k_\mathrm{BL}^\mathrm{opt}(f) + (N - 2 \ell_b(f)) k_\infty^\mathrm{opt}(f)$. At sufficiently large $N$, the interior contribution dominates, giving the leading bulk asymptote $k_\mathrm{chain}^\mathrm{opt}(N; f) \sim N k_\infty^\mathrm{opt}(f)$. Thus, $k_\infty^\mathrm{opt}(f)$ represents the thermodynamic-limit per-bond scission rate of the finite-bending chain. For $\beta k_\phi\pi^2 = 1.82 \times 10^3$, this regime is reached only around $N \sim 10^3$, far beyond the boundary-layer scale $\ell_b(f) = 13$. This separation of scales reflects the exponentially larger rate contributions from lower-barrier bonds within the boundary layers, with the terminal bonds contributing most strongly. 

\begin{figure}
    \includegraphics{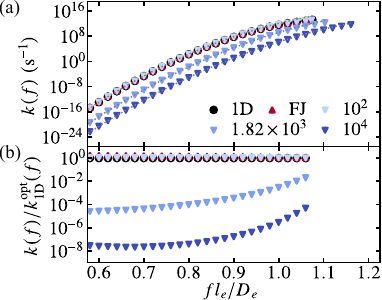}
    \caption{Force dependence of the thermodynamic-limit per-bond scission rate. (a) Absolute rates versus dimensionless force $f l_e/D_e$: $k_\infty^\mathrm{opt}(f)$ for finite bending stiffnesses $\beta k_\phi\pi^2 = 10^2$, $1.82 \times 10^3$, and $10^4$, $k_\mathrm{1D}^\mathrm{opt}(f)$ for the 1D reference, and $k_\mathrm{FJ}^\mathrm{opt}(f)$ for the FJ limit. Symbols show the full TST rates, with each series terminating at its chain-level critical force $f_c$. Dotted lines show selected local quadratic approximations centered at $f_\star l_e/D_e = 0.7$; the FJ and $\beta k_\phi\pi^2 = 10^2$ approximations are omitted because they nearly overlap with the 1D curve. (b) The same rates normalized by $k_\mathrm{1D}^\mathrm{opt}(f)$, with each series terminating at $f_\mathrm{cut} = \min\{f_c, f_c^\mathrm{1D}\}$.}
    \label{fig:force_dependence}
\end{figure}

We now turn to the force dependence of $k_\infty^\mathrm{opt}(f)$. Fig. \ref{fig:force_dependence} compares this rate for $\beta k_\phi\pi^2 = 10^2$, $1.82 \times 10^3$, and $10^4$ with $k_\mathrm{1D}^\mathrm{opt}(f)$ and $k_\mathrm{FJ}^\mathrm{opt}(f)$. Panel (a) focuses on the absolute rate scale for $f l_e/D_e \gtrsim 0.6$, with the full force range shown in Supplemental Material, Sec. S7.1 \cite{supplemental}. For the 1D reference, $k_\mathrm{1D}^\mathrm{opt}(f)$ rises from $1.10 \times 10^{-13} ~\mathrm{s}^{-1}$ at $f l_e/D_e = 0.6$ to $2.20 \times 10^{-5}~\mathrm{s}^{-1}$ at $f l_e/D_e = 0.7$, reducing the mean single-bond lifetime from $2.88 \times 10^5~\mathrm{yr}$ to $12.63~\mathrm{h}$. As $f$ approaches the critical force $f_c^\mathrm{1D} = a D_e/2 \approx 7.96~\mathrm{nN}$ ($f_c^\mathrm{1D} l_e / D_e = 1.075$), the rate approaches $1.18 \times 10^{13} ~\mathrm{s}^{-1}$. The other rates increase similarly with force but terminate at their respective chain-level critical forces. The FJ limit gives $f_c^\mathrm{FJ} l_e / D_e = 1.072$, slightly below the 1D value, consistent with its lower activation barrier discussed above. At finite bending stiffness, the terminal bonds set $f_c$, so the $k_\infty^\mathrm{opt}(f)$ series terminate while $\Delta \mathcal{W}_\infty(f)$ remains finite. For the three stiffnesses shown, the critical forces are $f_c l_e / D_e = 1.073$, $1.103$, and $1.161$, respectively, indicating that stronger bending constraints hinder force-induced barrier loss.

To isolate the 3D conformational effects from the overall force-induced rate increase, panel (b) shows the rates normalized by $k_\mathrm{1D}^\mathrm{opt}(f)$. The ratio $k_\mathrm{FJ}^\mathrm{opt}(f) / k_\mathrm{1D}^\mathrm{opt}(f)$ remains below $1.68$ over the range shown and decreases toward unity with increasing force, as force-induced alignment reduces the difference between the FJ and 1D descriptions. At lower forces, however, the difference can be much larger; at zero force, the FJ rate is $26.7$ times the 1D value. For $\beta k_\phi\pi^2 = 10^2$, the normalized rate remains only slightly below the FJ result. At stronger bending stiffness, these orientational constraints raise $\Delta \mathcal{W}_\infty(f)$ and substantially reducing $k_\infty^\mathrm{opt}(f)$ (Supplemental Material, Secs. S5.2 and S6.2 \cite{supplemental}). At $f l_e/D_e = 0.7$, the normalized rates are $4.10 \times 10^{-5}$ and $2.15 \times 10^{-8}$ for $\beta k_\phi\pi^2 = 1.82 \times 10^3$ and $10^4$, respectively. As the force increases further, the angular weights become confined to a narrower orientation range, while the direct force coupling increasingly governs barrier reduction (Secs. S5.3 and S6.3). Correspondingly, the normalized finite-bending rates increase with force. 

To obtain a simple local description of the force dependence, we denote the thermodynamic-limit per-bond scission rate generically by $k(f)$ and expand $\ln{k(f)}$ to second order about a reference force $f_\star$:
\begin{equation}
    \label{eq:simple_force_dependence}
    k(f) \approx k(f_\star) e^{\beta \lambda_\star (f - f_\star) - \frac{1}{2}\beta \kappa_\star (f - f_\star)^2},
\end{equation}
where $\lambda_\star = \beta^{-1} \left.{d \ln{k}/ d f}\right\vert_{f = f_\star}$ is the effective activation length and $\kappa_\star = -\beta^{-1} \left.{d^2 \ln{k} / d f^2}\right\vert_{f = f_\star}$ measures its force sensitivity (Supplemental Material, Sec. S7.2 \cite{supplemental}). This quadratic form parallels the extended Bell model \cite{konda2011chemical,makarov2016perspective}, but here the coefficients are evaluated at a finite $f_\star$ directly from the full TST rate in Eq. (\ref{eq:k_bond_with_W}), rather than from a low-force expansion about zero force. In the high-barrier regime, where the Arrhenius form in Eq. (\ref{eq:Arrhenius_form}) applies, neglecting the weaker force dependence of the prefactor gives $\lambda_\star \approx -d\Delta\mathcal{W}/df\vert_{f=f_\star}$. For the 1D reference, where force enters the PMF through the linear coupling $-fl$, this reduces to $\lambda_\star \approx l_\mathrm{1D}^\mathrm{b}(f_\star) - l_\mathrm{1D}^\mathrm{m}(f_\star)$, recovering the conventional activation-length interpretation \cite{makarov2016perspective}. In 3D, orientational contributions also enter the force dependence of the PMF barrier, so $\lambda_\star$ can no longer be identified solely with this geometric separation.

As shown in Fig. \ref{fig:force_dependence}(a), for $f_\star l_e/D_e = 0.7$, the quadratic approximation reproduces the full TST rates over $0.6 \leq f l_e/D_e \leq 0.8$ with maximum relative errors ranging from $1.7\%$ to $14.6\%$ across the 1D, FJ, and finite-bending cases (Supplemental Material, Table S1 \cite{supplemental}). At this reference force, $\lambda_\star$ varies only weakly, remaining between $0.63 l_e$ and $0.65 l_e$, whereas $\kappa_\star$ is reduced at larger bending stiffnesses by up to $22.7 \%$ relative to the 1D value. Thus, despite the orders-of-magnitude changes in the absolute rate, bending correlations only modestly modify the local force dependence.

In summary, we have developed a bond-resolved statistical-mechanical TST framework for chain scission in a 3D polymer chain under tension. In the FJ limit, the additional 3D configurational freedom yields higher rupture rates than the collinear 1D reference, with this enhancement diminishing as the force increases. Finite bending stiffness makes rupture bond-position dependent, with higher rates near the chain ends and a common interior rate, thereby producing finite-size effects in the chain-scission rate. At high stiffness, the FJ enhancement can be reversed, yielding rupture rates orders of magnitude below the 1D reference. The present PMF-based TST framework neglects dynamical corrections such as friction and recrossing. It also assumes constant-force equilibrium conditions. Incorporating such dynamical effects and extending the framework to nonequilibrium loading are natural extensions of the present work. Overall, this work provides molecularly resolved chain-scission rates as inputs for network-scale models of damage and fracture, laying a foundation for linking macroscopic polymer failure to the 3D conformational statistics of individual load-bearing chains.

\textit{Acknowledgments}---The authors thank Prof. David Manolopoulos from University of Oxford for important discussions and helpful comments on the manuscript. J.Z. acknowledges the support of an EPSRC DTP studentship at the University of Oxford [EP/W524311/1]. L.B. acknowledges supports from UKRI through a Future Leaders Fellowship [MR/W006995/1].

\textit{Data availability}---The data that support the findings of this Letter will be openly available \cite{Zhu617}.

\nocite{manca2012elasticity,buche2022freely,buche2021chain,khodayeki2022force}
\bibliography{bibfile.bib}

\end{document}


\title{Supplementary Material: Conformation-Mediated Kinetics of Polymer Chain Scission under Tension}
\author{Jie Zhu}
\affiliation{Department of Engineering Science, University of Oxford, Oxford OX1 3PJ, United Kingdom.}

\author{Laurence Brassart}
\affiliation{Department of Engineering Science, University of Oxford, Oxford OX1 3PJ, United Kingdom.}

\maketitle

\renewcommand{\theequation}{S\arabic{equation}}
\renewcommand{\thefigure}{S\arabic{figure}}
\renewcommand{\thetable}{S\arabic{table}}
\renewcommand{\bibnumfmt}[1]{[S#1]}
\renewcommand{\citenumfont}[1]{S#1}

\section{S1 Bond-resolved TST framework for chain scission}
\subsection{S1.1 Model system and Hamiltonian}
We consider a polymer chain composed of $N$ bonds connecting $N+1$ atoms, with positions $\{\vec{r}_0,\vec{r}_1,\ldots,\vec{r}_N\}$. One end of the chain is fixed at $\vec{r}_0=\vec{0}$, and a constant external force $\vec{f}$ is applied to the free end at $\vec{r}_N$ (Fig. 1b of the main text). All atoms are assumed to have the same mass $m$. Since the fixed endpoint is treated as a constraint, only atoms $1,\ldots,N$ are dynamical and have momenta $\{\vec{p}_1,\ldots,\vec{p}_N\}$. The corresponding augmented Hamiltonian is
\begin{equation}
    H(\{\vec{r}_i\}_{i=1}^N, \{\vec{p}_i\}_{i=1}^N; \vec{f}) = K(\{\vec{p}_i\}_{i=1}^N) + V(\{\vec{r}_i\}_{i=1}^N) - \vec{f}\cdot\vec{r}_{N},
\end{equation}
where 
\begin{equation}
    \label{seq:kinetic-energy}
    K(\{\vec{p}_i\}_{i=1}^N) = \sum_{i=1}^{N}\frac{\vec{p}_i\cdot\vec{p}_i}{2m}
\end{equation}
is the kinetic energy of the mobile atoms, $V(\{\vec{r}_i\}_{i=1}^N)$ is the internal interaction energy evaluated with the fixed endpoint, and $-\vec{f}\cdot\vec{r}_{N}$ is the potential associated with the applied force.

To simplify the configurational description, we introduce the bond vectors
\begin{equation}
    \vec{l}_i=\vec{r}_i-\vec{r}_{i-1}, \quad
    i=1,\ldots,N.
\end{equation} 
With $\vec{r}_0$ fixed, the chain configuration can be equivalently described by the set of bond vectors $(\vec{l}_1,\ldots,\vec{l}_N)$. Choosing the direction of the applied force as the polar axis, each bond vector can then be written in spherical coordinates as
\begin{equation}
    \vec{l}_i = l_i \left(\sin{\theta_i}\cos{\varphi_i},~\sin{\theta_i}\sin{\varphi_i},~\cos{\theta_i}\right),
\end{equation}
where $l_i = \|\vec{l}_i\|$ is the length of bond $i$, $\theta_i\in[0,\pi]$ is the polar angle with respect to this axis, and $\varphi_i\in[0,2\pi)$ is the azimuthal angle. The bond angle between neighboring bonds $\vec{l}_{i-1}$ and $\vec{l}_i$, for $i = 2,\ldots,N$, is
\begin{equation}
    \phi_i(\theta_i,\theta_{i-1},\omega_i) = \angle(\vec{l}_{i-1},\vec{l}_i) =  \arccos{\left[\sin{\theta_i}\sin{\theta_{i-1}}\cos{\omega_i}+\cos{\theta_i}\cos{\theta_{i-1}}\right]},
\end{equation}
where $\omega_i = \varphi_i-\varphi_{i-1}$ is the relative azimuthal angle. 

In terms of the bond vectors, $\vec{r}_{N} = \sum_{i=1}^N \vec{l}_i$, so the force contribution becomes
\begin{equation}
    - \vec{f}\cdot\vec{r}_{N} = -\sum_{i=1}^N \vec{f}\cdot\vec{l}_i
    = -\sum_{i=1}^N fl_i \cos{\theta_i},
\end{equation}
where $f = \|\vec{f}\|$ is the magnitude of the applied force. In the zero-force limit, no physical polar direction can be selected; the reference axis used to define $\theta_i$ is arbitrary, and the force contribution vanishes identically.

The internal interaction energy includes bond stretching and bond-angle deformation. Torsional interactions are neglected, corresponding to freely rotating dihedral degrees of freedom. Nonbonded intrachain interactions are also neglected. Thus,
\begin{equation}
    V(\{\vec{r}_i\}_{i=1}^N) \equiv V(\{\vec{l}_i\}_{i=1}^N) = \sum_{i=1}^{N}v_\mathrm{str}{(l_i)} + \sum_{i=2}^{N}v_\mathrm{ben}{(\phi_i)}.
\end{equation}
The stretching contribution is described by a Morse potential,
\begin{equation}
    v_\mathrm{str}(l) = D_e\left[1 - e^{-a(l-l_e)}\right]^2,
\end{equation}
where $D_e$ is the bond dissociation energy, $a$ sets the inverse range of the Morse well, and $l_e$ is the equilibrium bond length. This form captures bond anharmonicity and finite dissociation energy.

The bond-angle contribution penalizes deviations from the equilibrium bond angle and is modeled as
\begin{equation}
    v_\mathrm{ben}(\phi) = \frac{1}{2}k_{\phi}\left(\phi-\phi_e\right)^2,
\end{equation}
where $k_\phi$ is the bending stiffness and $\phi_e$ is the equilibrium bond angle. This term provides a minimal description of nearest-neighbor orientational correlations. 

Combining the internal interaction energy with the force contribution, the Hamiltonian can equivalently be written in bond-vector variables as
\begin{equation}
    H(\{\vec{l}_i\}_{i=1}^N, \{\vec{p}_i\}_{i=1}^N; f) = K(\{\vec{p}_i\}_{i=1}^N) + U(\{l_i,\theta_i,\varphi_i\}_{i=1}^N;f),
\end{equation}
where the configurational energy is
\begin{equation}
    U(\{l_i,\theta_i,\varphi_i\}_{i=1}^N;f) = \sum_{i=1}^{N}v_\mathrm{str}{(l_i)} + \sum_{i=2}^{N}v_\mathrm{ben}{(\phi_i)} - \sum_{i=1}^N fl_i \cos{\theta_i}.
\end{equation}
With this choice of coordinates, the configurational energy depends on the external force only through the scalar magnitude $f$. Moreover, with $\vec{r}_0$ fixed, the transformation from $(\vec{r}_1,\ldots,\vec{r}_N)$ to $(\vec{l}_1,\ldots,\vec{l}_N)$ has unit Jacobian. Keeping the momenta in the atomic representation, we write the phase-space measure as
\begin{equation}
    d\Gamma =  \prod_{i=1}^N \,d\vec{p}_i\,d\vec{l}_i = \prod_{i=1}^N \,d\vec{p}_i l_i^2\sin{\theta_i}\,dl_i\,d\theta_i\,d\varphi_i.
\end{equation}

\subsection{S1.2 Multi-channel rupture and TST formulation}
We formulate chain scission as a multi-channel first-rupture problem in bond-length space. Each bond $i$ is assigned a rupture threshold $l_i^\ddagger$ and is regarded as ruptured when $l_i \geq l_i^\ddagger$. We collect these thresholds as $\boldsymbol{l}^\ddagger = \{l_i^\ddagger\}_{i=1}^N$. The chain remains intact only when all bonds lie below their respective thresholds, defining the intact basin
\begin{equation}
    \mathcal{R}(\boldsymbol{l}^\ddagger) = \{l_i < l_i^\ddagger,~\forall i = 1,\ldots,N\}.
\end{equation}
Chain scission is defined by the rupture of any bond, so the dissociated region is
\begin{equation}
    \mathcal{P}(\boldsymbol{l}^\ddagger) = \{\exists i \text{ such that } l_i \geq l_i^\ddagger\}.
\end{equation} 
The dividing surface between these regions is the boundary of the intact basin,
\begin{equation}
    \Sigma(\boldsymbol{l}^\ddagger) = \partial \mathcal{R}(\boldsymbol{l}^\ddagger) = \bigcup_{i= 1}^N \{l_i = l_i^\ddagger,~l_j \leq l_j^\ddagger~\text{for all}~j \neq i\}.
\end{equation}

Within transition-state theory (TST), the chain-scission rate is estimated as the positive equilibrium flux through $\Sigma(\boldsymbol{l}^\ddagger)$ from the intact basin to the dissociated region, normalized by the equilibrium population of the intact basin. We write this rate as
\begin{equation}
    k_\mathrm{chain}(\boldsymbol{l}^\ddagger;f) = \frac{\Phi_\Sigma(\boldsymbol{l}^\ddagger;f)}{Z_\mathcal{R}(\boldsymbol{l}^\ddagger;f)}.
\end{equation}
At this stage, $\boldsymbol{l}^\ddagger$ is an arbitrary but fixed threshold set. Different choices of $\boldsymbol{l}^\ddagger$ define different TST dividing surfaces and hence different TST rate estimates. Variational TST can then be used to select the threshold set that minimizes the resulting rate \cite{truhlar1980variational,makarov2015single}.

The intact-basin population is represented by the restricted partition function,
\begin{equation}
    Z_\mathcal{R}(\boldsymbol{l}^\ddagger;f) =  \int d\Gamma\,e^{-\beta H(\{\vec{l}_k\}_{k=1}^N, \{\vec{p}_k\}_{k=1}^N; f)} \prod_{i=1}^N \Theta(l_i^\ddagger - l_i),
\end{equation}
where $\Theta(x)$ is the Heaviside step function, equal to unity for $x>0$ and zero otherwise, and $\beta \equiv (k_BT)^{-1}$, with $k_B$ the Boltzmann constant and $T$ the temperature.

The dividing surface contains $N$ bond-resolved rupture channels,
\begin{equation}
    \Sigma_i(\boldsymbol{l}^\ddagger) = \{l_i = l_i^\ddagger,~l_j < l_j^\ddagger~\text{for all}~j \neq i\}.
\end{equation}
Each channel $\Sigma_i$ contains configurations in which bond $i$ is the unique bond at its rupture threshold while all other bonds remain intact. The remaining parts of $\Sigma$, where two or more bonds are simultaneously at their thresholds, satisfy two or more independent threshold constraints and are therefore lower-dimensional than the individual channel surfaces. Because the TST flux is obtained by integrating over the channel surfaces, these lower-dimensional sets occupy zero surface measure and make no contribution to the total positive flux. The flux thus decomposes additively as
\begin{equation}
    \Phi_\Sigma(\boldsymbol{l}^\ddagger;f) = \sum_{i=1}^N \Phi_i(\boldsymbol{l}^\ddagger;f).
\end{equation}
The equilibrium flux through channel $i$ is
\begin{equation}
    \Phi_i(\boldsymbol{l}^\ddagger;f) = \int d\Gamma\,e^{-\beta H(\{\vec{l}_k\}_{k=1}^N, \{\vec{p}_k\}_{k=1}^N; f)} \delta(l_i -l_i^\ddagger) \Theta(\dot{l}_i) \dot{l}_i \prod_{j \neq i} \Theta(l_j^\ddagger - l_j),
\end{equation}
where $\delta(\cdot)$ is the Dirac delta function, and $\dot{l}_i$ is the instantaneous rate of change of bond length $l_i$, i.e., the normal velocity through the channel surface $\Sigma_i$. The factor $\Theta(\dot{l}_i)\dot{l}_i$ selects positive crossings from the intact basin to the dissociated region.

Since all channels are normalized by the same intact-basin population, we define the bond-resolved TST rate as 
\begin{equation}
    k_{\mathrm{bond},i}(\boldsymbol{l}^\ddagger;f) = \frac{\Phi_i(\boldsymbol{l}^\ddagger;f)}{Z_\mathcal{R}(\boldsymbol{l}^\ddagger;f)}.
\end{equation}
The additive flux decomposition then gives
\begin{equation}
    k_\mathrm{chain}(\boldsymbol{l}^\ddagger;f) = \sum_{i=1}^N k_{\mathrm{bond},i}(\boldsymbol{l}^\ddagger;f).
\end{equation}

\subsection{S1.3 Factorization into kinetic and configurational contributions}
Using the decomposition $H = K + U$ and the phase-space measure defined in Sec. S1.1, the Boltzmann weight factorizes as
\begin{equation}
    e^{-\beta H(\{\vec{l}_k\}_{k=1}^N, \{\vec{p}_k\}_{k=1}^N; f)} = e^{-\beta K(\{\vec{p}_k\}_{k=1}^N)} e^{-\beta U(\{l_k,\theta_k,\varphi_k\}_{k=1}^N;f)}.
\end{equation}
The phase-space measure separates into momentum and configurational parts,
\begin{equation}
    d\Gamma = d\Gamma_\mathrm{mom}\,d\Gamma_\mathrm{conf},
\end{equation}
with
\begin{equation}
    d\Gamma_\mathrm{mom} =  \prod_{k=1}^N \,d\vec{p}_k, \quad
    d\Gamma_\mathrm{conf} = \prod_{k=1}^N l_k^2\sin{\theta_k}\,dl_k\,d\theta_k\,d\varphi_k.
\end{equation}

The intact-basin partition function therefore factorizes as
\begin{equation}
    Z_\mathcal{R}(\boldsymbol{l}^\ddagger;f) = \left(\int d\Gamma_\mathrm{mom}\,e^{-\beta K(\{\vec{p}_k\}_{k=1}^N)}\right) Z_\mathcal{R}^\mathrm{conf}(\boldsymbol{l}^\ddagger;f),
\end{equation}
where
\begin{equation}
    Z_\mathcal{R}^\mathrm{conf}(\boldsymbol{l}^\ddagger;f) = \int d\Gamma_\mathrm{conf}\,e^{-\beta U(\{l_k,\theta_k,\varphi_k\}_{k=1}^N;f)} \prod_{i=1}^N \Theta(l_i^\ddagger - l_i)
\end{equation}
is the configurational partition function restricted to the intact basin.

Similarly, the bond-resolved flux through channel $i$ separates into a momentum factor and a configurational surface factor,
\begin{equation}
    \Phi_i(\boldsymbol{l}^\ddagger;f) = \left(\int d\Gamma_\mathrm{mom}\,e^{-\beta K(\{\vec{p}_k\}_{k=1}^N)} \Theta(\dot{l}_i) \dot{l}_i\right) \Phi_i^\mathrm{conf}(\boldsymbol{l}^\ddagger;f),
\end{equation}
where 
\begin{equation}
    \Phi_i^\mathrm{conf}(\boldsymbol{l}^\ddagger;f) = \int d\Gamma_\mathrm{conf}\,e^{-\beta U(\{l_k,\theta_k,\varphi_k\}_{k=1}^N;f)} \delta(l_i -l_i^\ddagger) \prod_{j \neq i} \Theta(l_j^\ddagger - l_j)
\end{equation}
is the configurational surface factor for channel $i$. For a fixed configuration, $\dot{l}_i$ contains projections of the atomic momenta along the instantaneous bond direction $\hat{l}_i = \vec{l}_i/l_i$. However, the Gaussian momentum distribution generated by $K$ is isotropic. Therefore, the momentum average of $\Theta(\dot{l}_i) \dot{l}_i$ depends only on the variance of the projected relative velocity, not on the specific orientation of $\hat{l}_i$. This justifies separating the momentum factor from the configurational integral.

Introducing the normalized momentum average
\begin{equation}
    \langle A \rangle_\mathrm{mom} \equiv \frac{\displaystyle \int d\Gamma_\mathrm{mom}\,e^{-\beta K(\{\vec{p}_k\}_{k=1}^N)} A}{\displaystyle \int d\Gamma_\mathrm{mom}\,e^{-\beta K(\{\vec{p}_k\}_{k=1}^N)}},
\end{equation}
the bond-resolved TST rate becomes
\begin{equation}
    \label{seq:bond-resolved-rate-original}
    k_{\mathrm{bond},i}(\boldsymbol{l}^\ddagger;f) = \langle \Theta(\dot{l}_i) \dot{l}_i \rangle_\mathrm{mom} \frac{\Phi_i^\mathrm{conf}(\boldsymbol{l}^\ddagger;f)}{Z_\mathcal{R}^\mathrm{conf}(\boldsymbol{l}^\ddagger;f)}.
\end{equation}
This expression separates the kinetic prefactor from the configurational contribution.

We now evaluate the momentum average explicitly. The time derivative of a bond vector is
\begin{equation}
    \dot{\vec{l}}_i = \dot{\vec{r}}_i - \dot{\vec{r}}_{i-1} = \frac{\vec{p}_i}{m} - \frac{\vec{p}_{i-1}}{m},
\end{equation}
with the convention $\vec{p}_0 = \vec{0}$ for the fixed end. Hence the bond-length velocity is
\begin{equation}
    \dot{l}_i = \hat{l}_i \cdot \dot{\vec{l}}_i = \frac{1}{m} (\hat{l}_i \cdot \vec{p}_i - \hat{l}_i \cdot \vec{p}_{i-1}).
\end{equation}
For the first bond,
\begin{equation}
    \dot{l}_1 = \hat{l}_1 \cdot \dot{\vec{l}}_1 = \frac{1}{m} \hat{l}_1 \cdot \vec{p}_1.
\end{equation}
Since the kinetic energy is quadratic (Eq. \ref{seq:kinetic-energy}), the Cartesian components of each $\vec{p}_k$ are independent Gaussian variables with zero mean and variance $m/\beta$. Equivalently,
\begin{equation}
     \langle p_{k,\alpha} p_{k,\beta} \rangle_\mathrm{mom} = \frac{m}{\beta} \delta_{\alpha\beta}.
\end{equation}
By rotational invariance, the projection of $\vec{p}_k$ along any unit vector has the same variance,
\begin{equation}
    \langle (\hat{l}_i \cdot \vec{p}_{k})^2 \rangle_\mathrm{mom} = \frac{m}{\beta}.
\end{equation}
It follows that for the first bond,
\begin{equation}
    \langle \dot{l}_1^2 \rangle_\mathrm{mom} = \frac{1}{m^2} \langle (\hat{l}_1 \cdot \vec{p}_1)^2 \rangle_\mathrm{mom} = \frac{1}{m\beta}.
\end{equation}
For any bond with $i \geq 2$, the two random variables $\hat{l}_i \cdot \vec{p}_i$ and $\hat{l}_i \cdot \vec{p}_{i-1}$ are independent Gaussians with variance $m/\beta$. Thus
\begin{equation}
    \langle \dot{l}_i^2 \rangle_\mathrm{mom} = \frac{1}{m^2} \langle (\hat{l}_i \cdot \vec{p}_i - \hat{l}_i \cdot \vec{p}_{i-1})^2 \rangle_\mathrm{mom} = \frac{2}{m\beta}, \quad i \geq 2.
\end{equation}
It is convenient to define an effective mass $m_i^l$ for the bond-length degree of freedom through
\begin{equation}
    \langle \dot{l}_i^2 \rangle_{\mathrm{mom}} = \frac{1}{m^l_i\beta}.
\end{equation}
This gives
\begin{equation}
    m^l_1 = m; \quad m^l_i = \frac{m}{2}, \quad i \geq 2.
\end{equation}
Thus, the first bond behaves as a mass $m$ attached to a fixed wall, whereas each bond with $i \geq 2$ corresponds to the relative coordinate of two equal masses and therefore carries the reduced mass $m/2$.

Since $\dot{l}_i$ is a one-dimensional Gaussian variable with zero mean and variance $1/(m_i^l\beta)$, its positive one-sided average is
\begin{equation}
    \langle \Theta(\dot{l}_i) \dot{l}_i \rangle_{\mathrm{mom}} = \frac{1}{\sqrt{2\pi m_i^l\beta}}.
\end{equation}
This is the kinetic prefactor entering the TST rate for bond $i$. Apart from the fixed-end bond, all bonds have the same kinetic prefactor because $m^l_i = m/2$ for $i \geq 2$. Therefore, bond-to-bond variations among bonds $i \geq 2$ arise from configurational statistics alone. In the next subsection, we analyze this configurational contribution in terms of a potential of mean force (PMF) along the bond-length coordinate.

\subsection{S1.4 PMF representation and self-consistent optimal bond-resolved TST rate}
Following the factorization derived in Sec. S1.3, we now focus on the configurational contribution to the bond-resolved TST rate. For a given bond $i$, we introduce the constrained configurational integral
\begin{equation}
    \mathcal{G}_i(l;\boldsymbol{l}_{-i}^\ddagger,f) = \int d\Gamma_\mathrm{conf}\,e^{-\beta U(\{l_k,\theta_k,\varphi_k\}_{k=1}^N;f)} \delta(l_i -l) \prod_{j \neq i} \Theta(l_j^\ddagger - l_j),
\end{equation}
where $\boldsymbol{l}_{-i}^\ddagger = \{l_j^\ddagger\}_{j \neq i}$. This quantity gives the configurational weight at fixed bond length $l_i=l$, while all other bonds remain below their rupture thresholds. Integrating over the allowed interval of bond $i$ yields the configurational intact-basin partition function,
\begin{equation}
    \label{Sqe:Z_Rconf}
    Z_\mathcal{R}^\mathrm{conf}(\boldsymbol{l}^\ddagger; f) = \int_0^{l_i^\ddagger} \mathcal{G}_i(l; \boldsymbol{l}_{-i}^\ddagger, f)\,dl,
\end{equation}
and evaluating the same constrained integral at $l = l_i^\ddagger$ gives the configurational surface factor for channel $i$,
\begin{equation}
    \Phi_i^\mathrm{conf}(\boldsymbol{l}^\ddagger; f) = \mathcal{G}_i(l_i^\ddagger; \boldsymbol{l}_{-i}^\ddagger, f).
\end{equation}

We then define the one-dimensional potential of mean force (PMF) along the length of bond $i$ as
\begin{equation}
    \label{Seq:PMF_def}
    \mathcal{W}_i(l; \boldsymbol{l}_{-i}^\ddagger, f) = -{\beta}^{-1}\ln{\mathcal{G}_i(l; \boldsymbol{l}_{-i}^\ddagger, f)},
\end{equation}
so that 
\begin{equation}
    \mathcal{G}_i(l; \boldsymbol{l}_{-i}^\ddagger, f) = e^{-\beta \mathcal{W}_i(l; \boldsymbol{l}_{-i}^\ddagger, f)}.
\end{equation}
The PMF represents the effective free-energy profile of bond $i$ along its length coordinate, with all other bonds constrained below their rupture thresholds.

Using this definition, the bond-resolved TST rate (Eq. \ref{seq:bond-resolved-rate-original}) becomes
\begin{equation}
    k_{\mathrm{bond},i}(\boldsymbol{l}^\ddagger;f) = \frac{1}{\sqrt{2\pi m_i^l \beta}} \frac{\displaystyle e^{-\beta \mathcal{W}_i(l_i^\ddagger; \boldsymbol{l}_{-i}^\ddagger, f)}}{\displaystyle \int_0^{l_i^\ddagger} e^{-\beta \mathcal{W}_i(l; \boldsymbol{l}_{-i}^\ddagger, f)}\,dl}.
\end{equation}
This form shows that the configurational contribution to the rate is fully determined by the PMF. Any additive constant in $\mathcal{W}_i(l; \boldsymbol{l}_{-i}^\ddagger, f)$ that is independent of $l$ cancels in the ratio, so only the $l$-dependence of the PMF is relevant.

For moderate forces, $\mathcal{W}_i(l; \boldsymbol{l}_{-i}^\ddagger, f)$ typically exhibits a metastable bonded well with a local minimum at $l_i^\mathrm{m}(\boldsymbol{l}_{-i}^\ddagger, f)$ and a barrier top at $l_i^\mathrm{b}(\boldsymbol{l}_{-i}^\ddagger, f)$. For a prescribed $\boldsymbol{l}_{-i}^\ddagger$, these points satisfy
\begin{eqnarray}
    \label{seq:stationarity_and_curvature_condition}
    \mathcal{W}'_i(l_i^\mathrm{m}; \boldsymbol{l}_{-i}^\ddagger, f) &=& 0, \quad \mathcal{W}''_i(l_i^\mathrm{m}; \boldsymbol{l}_{-i}^\ddagger, f) > 0; \\ \nonumber
    \mathcal{W}'_i(l_i^\mathrm{b}; \boldsymbol{l}_{-i}^\ddagger, f) &=& 0, \quad \mathcal{W}''_i(l_i^\mathrm{b}; \boldsymbol{l}_{-i}^\ddagger, f) < 0,
\end{eqnarray}
where primes denote derivatives with respect to the bond length $l$. The minimum $l_i^\mathrm{m}(\boldsymbol{l}_{-i}^\ddagger, f)$ corresponds to the most probable length of bond $i$ under the prescribed intact constraints. Motivated by the variational-TST framework \cite{truhlar1980variational,makarov2015single}, we use the barrier top $l_i^\mathrm{b}(\boldsymbol{l}_{-i}^\ddagger, f)$ as a PMF-based threshold for rupture channel $i$. Because the barrier location depends on the prescribed thresholds $\boldsymbol{l}_{-i}^\ddagger$, we determine the threshold set self-consistently by requiring each threshold to coincide with the corresponding PMF barrier,
\begin{equation}
    l_{i,\mathrm{opt}}^\ddagger(f) = l_i^\mathrm{b}(\boldsymbol{l}_{-i,\mathrm{opt}}^\ddagger(f), f), \quad i = 1,\ldots, N.
\end{equation}
After convergence, we write the resulting PMF as $\mathcal{W}_i(l;f)$, its minimum as $l_i^\mathrm{m}(f)$, and its barrier location as $l_i^\mathrm{b}(f)$, with $l_{i,\mathrm{opt}}^\ddagger(f) = l_i^\mathrm{b}(f)$. The corresponding optimal bond-resolved TST rate is
\begin{equation}
\label{Sqe:k_bond_with_W}
    k_{\mathrm{bond},i}^\mathrm{opt}(f) = \frac{1}{\sqrt{2\pi m_i^l \beta}} \frac{\displaystyle e^{-\beta \mathcal{W}_i(l_i^\mathrm{b}(f);f)}}{\displaystyle \int_0^{l_i^\mathrm{b}(f)} e^{-\beta \mathcal{W}_i(l;f)}\,dl}.
\end{equation}

As the applied force increases, the minimum $l_i^\mathrm{m}(f)$ and the barrier $l_i^\mathrm{b}(f)$ approach each other and eventually merge. This occurs at a bond-dependent critical force $f_i^c$, defined by
\begin{equation}
    \mathcal{W}'_i(l_i^c;f_i^c) = 0, \quad \mathcal{W}''_i(l_i^c;f_i^c) = 0.
\end{equation}
Here $l_i^c$ is the critical bond length at which the metastable minimum and barrier top coincide. The chain-level critical force is
\begin{equation}
    f_c = \min_{1 \leq i \leq N} f_i^c.
\end{equation}
For $f \geq f_c$, at least one bond no longer has a metastable bonded basin in its PMF, and rupture along that coordinate becomes barrierless. The present PMF-based TST construction is therefore applicable only for $f < f_c$. As $f \rightarrow f_c^-$, the barrier for the limiting bond becomes shallow and dynamical recrossings become significant, so the TST approximation breaks down.

When the bonded well is deep and well separated from the barrier, the PMF near its minimum $l_i^\mathrm{m}(f)$ can be approximated as
\begin{equation}
    \mathcal{W}_i(l;f) \approx \mathcal{W}_i(l_i^\mathrm{m}(f);f) + \frac{1}{2}m_i^l[\omega_i^\mathrm{m}(f)]^2 [l - l_i^\mathrm{m}(f)]^2,
\end{equation}
with
\begin{equation}
    m_i^l[\omega_i^\mathrm{m}(f)]^2 = \left.\frac{\partial^2 \mathcal{W}_i(l;f)}{\partial l^2}\right\vert_{l = l_i^\mathrm{m}(f)}.
\end{equation}
Under this approximation,
\begin{equation}
    \int_0^{l_i^\mathrm{b}(f)} e^{-\beta \mathcal{W}_i(l;f)}\,dl \approx e^{-\beta \mathcal{W}_i(l_i^\mathrm{m}(f);f)} \int_{-\infty}^\infty e^{-\frac{1}{2}\beta m_i^l[\omega_i^\mathrm{m}(f)]^2 (l - l_i^\mathrm{m}(f))^2}\,dl,
\end{equation}
which gives
\begin{equation}
    \int_0^{l_i^\mathrm{b}(f)} e^{-\beta \mathcal{W}_i(l;f)}\,dl \approx  \sqrt{\frac{2\pi}{\beta m_i^l[\omega_i^\mathrm{m}(f)]^2}} e^{-\beta \mathcal{W}_i(l_i^\mathrm{m}(f);f)}.
\end{equation}
Substituting this result into the optimal bond-resolved TST expression (Eq. \ref{Sqe:k_bond_with_W}) gives the Arrhenius form \cite{hanggi1990reaction}
\begin{equation}
    \label{Seq:Arrhenius_form}
    k_{\mathrm{bond},i}^\mathrm{opt}(f) \approx \frac{\omega_i^\mathrm{m}(f)}{2\pi} e^{-\beta \Delta \mathcal{W}_i(f)}, 
\end{equation}
where 
\begin{equation}
    \Delta \mathcal{W}_i(f) = \mathcal{W}_i(l_i^\mathrm{b}(f);f) - \mathcal{W}_i(l_i^\mathrm{m}(f);f)
\end{equation}
is the PMF barrier height. The prefactor ${\omega_i^\mathrm{m}(f)}/{2\pi}$, equal to the inverse vibrational period in the bonded well, represents the attempt frequency for bond $i$ to cross the barrier. Finally, summing over all rupture channels gives the optimal chain-scission rate,
\begin{equation}
    k_\mathrm{chain}^\mathrm{opt}(f) = \sum_{i=1}^N k_{\mathrm{bond},i}^\mathrm{opt}(f).
\end{equation}

\section{S2 Transfer-matrix approach for bond-length PMFs}
\subsection{S2.1 Exact transfer-matrix formulation}
In Sec. S1.4, the bond-resolved TST rate is expressed in terms of the constrained configurational integrals $\mathcal{G}_i(l; \boldsymbol{l}_{-i}^\ddagger, f)$ and the associated PMFs $\mathcal{W}_i(l; \boldsymbol{l}_{-i}^\ddagger, f)$. In this section, we develop an efficient transfer-matrix (TM) formulation for evaluating these quantities, building on our previous TM formulations for chain elasticity \cite{zhu2025stretching,zhu2026elasticity}. The basic idea is to reorganize the constrained configurational integral so that the local bond-length contribution is treated separately, while the nearest-neighbor orientational correlations are propagated through angular kernels.

Following the bond-dependent threshold formulation of Sec. S1.2, we first define the local intact weight associated with bond $i$ at fixed polar angle $\theta$ by integrating over the allowed bond-length interval:
\begin{equation}
    I_i(\theta; l_i^\ddagger, f) = \sin{\theta} \int_0^{l_i^\ddagger} e^{-\beta [v_\mathrm{str}(l) - fl\cos{\theta}]}l^2\,dl.
\end{equation}
Here the factor $l^2\sin\theta$ is inherited from the configurational measure introduced in Sec. S1.1.

The bending energy couples only neighboring bond orientations. For two successive bonds $i$ and $i+1$, with polar angles $\theta'$ and $\theta$, respectively, we integrate out the relative azimuthal angle $\omega$ and define the angular coupling kernel
\begin{equation}
    Q(\theta,\theta') = \int_0^{2\pi} e^{-\beta v_{\mathrm{ben}}(\phi(\theta,\theta',\omega))}\,d\omega.
\end{equation}

With $I_i$ and $Q$ defined, the constrained configurational integrals can be assembled through forward and backward messages. We denote the rupture thresholds to the left and right of bond $i$ by
\begin{equation}
    \boldsymbol{l}_{<i}^\ddagger = \{l_1^\ddagger,\ldots,l_{i-1}^\ddagger\}, 
    \quad \boldsymbol{l}_{>i}^\ddagger = \{l_{i+1}^\ddagger,\ldots,l_N^\ddagger\}.
\end{equation}
The forward message $L_i(\theta; \boldsymbol{l}_{<i}^\ddagger, f)$ collects only the intact-basin contribution of the subchain to the left of bond $i$, conditioned on bond $i$ having polar angle $\theta_i = \theta$. Since there is no bond to the left of bond $1$, the left boundary condition is
\begin{equation}
    L_1(\theta; \varnothing, f) = 1.
\end{equation}
For $i = 1, \ldots, N-1$, forward propagation is obtained by integrating over the polar angle of bond $i$, including the local intact weight of bond $i$ and the angular coupling between bonds $i$ and $i+1$:
\begin{equation}
    L_{i+1}(\theta; \boldsymbol{l}_{<i+1}^\ddagger, f) = \int_0^\pi L_i(\theta'; \boldsymbol{l}_{<i}^\ddagger, f) I_i(\theta'; l_i^\ddagger, f) Q(\theta,\theta') \,d\theta'.
\end{equation}

Similarly, the backward message $R_i(\theta; \boldsymbol{l}_{>i}^\ddagger, f)$ collects only the intact-basin contribution of the subchain to the right of bond $i$, conditioned on bond $i$ having polar angle $\theta_i = \theta$. Since there is no bond to the right of bond $N$, the right boundary condition is
\begin{equation}
    R_N(\theta; \varnothing, f) = 1.
\end{equation}
For $i = 2, \ldots, N$, backward propagation takes the form
\begin{equation}
    R_{i-1}(\theta; \boldsymbol{l}_{>i-1}^\ddagger, f) = \int_0^\pi Q(\theta',\theta) I_i(\theta'; l_i^\ddagger, f) R_i(\theta'; \boldsymbol{l}_{>i}^\ddagger, f) \,d\theta'.
\end{equation}
Here the integration variable $\theta'$ is the polar angle of bond $i$, while $\theta$ is the polar angle of bond $i-1$.

We next construct the constrained configurational integral $\mathcal{G}_i(l; \boldsymbol{l}_{-i}^\ddagger, f)$, in which bond $i$ is fixed at length $l$ while all other bonds remain below their own thresholds. At fixed $l$ and $\theta$, the local contribution of bond $i$ is
\begin{equation}
    \label{Seq:J_l_theta}
    J(l,\theta; f) = e^{-\beta [v_\mathrm{str}(l) - fl\cos{\theta}]}l^2 \sin{\theta}.
\end{equation}
Combining this fixed-length local contribution with the left and right messages gives
\begin{equation}
    \label{Seq:G_TM_original}
    \mathcal{G}_i(l; \boldsymbol{l}_{-i}^\ddagger, f) = 2\pi \int_0^\pi L_i(\theta; \boldsymbol{l}_{<i}^\ddagger, f) J(l, \theta; f) R_i(\theta; \boldsymbol{l}_{>i}^\ddagger, f)\,d\theta.
\end{equation}
The prefactor $2\pi$ comes from the free global azimuthal angle. Since $\boldsymbol{l}_{-i}^\ddagger = \boldsymbol{l}_{<i}^\ddagger \cup \boldsymbol{l}_{>i}^\ddagger$, the threshold dependence of $\mathcal{G}_i(l;\boldsymbol{l}_{-i}^\ddagger,f)$ is carried by the left and right messages.

Substituting the TM expression for $\mathcal{G}_i(l;\boldsymbol{l}_{-i}^\ddagger,f)$ into Eq. (\ref{Sqe:Z_Rconf}) gives the intact-basin configurational partition function
\begin{equation}
    \label{Seq:ZRconf_original}
    Z_\mathcal{R}^\mathrm{conf}(\boldsymbol{l}^\ddagger; f) = 2\pi \int_0^\pi L_i(\theta; \boldsymbol{l}_{<i}^\ddagger, f) I_i(\theta; l_i^\ddagger, f) R_i(\theta; \boldsymbol{l}_{>i}^\ddagger, f)\,d\theta.
\end{equation}
The product $L_i I_i R_i$ represents the full intact-basin weight conditioned on bond $i$ having polar angle $\theta_i=\theta$. The fact that $Z_\mathcal{R}^\mathrm{conf}$ does not depend on the chosen bond index $i$ provides an internal consistency check for the TM construction.

The associated PMF follows directly from its definition, Eq. (\ref{Seq:PMF_def}), in Sec. S1.4. This completes the exact TM formulation for evaluating the constrained configurational integrals and their associated PMFs. The next subsection develops a rescaled-message implementation for computing the PMFs and the self-consistent bond-dependent thresholds.

\subsection{S2.2 Numerically stable implementation via rescaled messages}
The exact TM formulation derived in Sec. S2.1 is not, in its raw form, numerically stable for long chains. Direct propagation of the messages $L_i(\theta; \boldsymbol{l}_{<i}^\ddagger, f)$ and $R_i(\theta; \boldsymbol{l}_{>i}^\ddagger, f)$ causes their overall magnitudes to grow or decay roughly exponentially with chain length, even though the physically relevant information is contained in their angular shapes. To improve numerical stability, we separate out the accumulated magnitudes from the angular dependence of the messages.

We write the forward and backward messages in the factorized form
\begin{equation}
    L_i(\theta; \boldsymbol{l}_{<i}^\ddagger, f) = A_i^L(\boldsymbol{l}_{<i}^\ddagger, f) P_i^L(\theta; \boldsymbol{l}_{<i}^\ddagger, f), \quad
    R_i(\theta; \boldsymbol{l}_{>i}^\ddagger, f) = A_i^R(\boldsymbol{l}_{>i}^\ddagger, f) P_i^R(\theta; \boldsymbol{l}_{>i}^\ddagger, f),
\end{equation}
where
\begin{equation}
    \int_0^\pi P_i^L(\theta; \boldsymbol{l}_{<i}^\ddagger, f)\,d\theta = 1, \quad
    \int_0^\pi P_i^R(\theta; \boldsymbol{l}_{>i}^\ddagger, f)\,d\theta = 1.
\end{equation}
Thus, $P_i^L$ and $P_i^R$ describe the normalized angular shapes of the forward and backward messages, while $A_i^L$ and $A_i^R$ carry their accumulated magnitudes.

From the boundary conditions $L_1(\theta;\varnothing,f)=1$ and $R_N(\theta;\varnothing,f)=1$, we obtain
\begin{eqnarray}
    A_1^L(\varnothing, f) &=& \pi, \quad
    P_1^L(\theta; \varnothing, f) = \frac{1}{\pi};\\ \nonumber
    A_N^R(\varnothing, f) &=& \pi, \quad
    P_N^R(\theta; \varnothing, f) = \frac{1}{\pi}.
\end{eqnarray}

To propagate the messages stably, we define the rescaled propagated messages by removing the accumulated magnitude from the previous step:
\begin{eqnarray}
    \hat{L}_{i+1}(\theta; \boldsymbol{l}_{<i+1}^\ddagger, f) &=& \frac{L_{i+1}(\theta; \boldsymbol{l}_{<i+1}^\ddagger, f)}{A_i^L(\boldsymbol{l}_{<i}^\ddagger, f)}, \quad i = 1,\ldots,N-1,\\ \nonumber
    \hat{R}_{i-1}(\theta; \boldsymbol{l}_{>i-1}^\ddagger, f) &=& \frac{R_{i-1}(\theta; \boldsymbol{l}_{>i-1}^\ddagger, f)}{A_i^R(\boldsymbol{l}_{>i}^\ddagger, f)}, \quad i = 2,\ldots,N.
\end{eqnarray}
Substituting the exact message recursions from Sec. S2.1 gives the stable propagation formulas
\begin{eqnarray}
    \hat{L}_{i+1}(\theta; \boldsymbol{l}_{<i+1}^\ddagger, f) &=& \int_0^\pi P^L_i(\theta'; \boldsymbol{l}_{<i}^\ddagger, f) I_i(\theta'; l_i^\ddagger, f) Q(\theta,\theta') \,d\theta',\\ \nonumber
    \hat{R}_{i-1}(\theta; \boldsymbol{l}_{>i-1}^\ddagger, f) &=& \int_0^\pi Q(\theta',\theta) I_i(\theta'; l_i^\ddagger, f) P^R_i(\theta'; \boldsymbol{l}_{>i}^\ddagger, f) \,d\theta'.
\end{eqnarray}
The norms of the rescaled propagated messages are
\begin{equation}
    \hat{A}_{i+1}^L(\boldsymbol{l}_{<i+1}^\ddagger, f) = \int_0^\pi \hat{L}_{i+1}(\theta; \boldsymbol{l}_{<i+1}^\ddagger, f)\,d\theta, \quad
    \hat{A}_{i-1}^R(\boldsymbol{l}_{>i-1}^\ddagger, f) = \int_0^\pi \hat{R}_{i-1}(\theta; \boldsymbol{l}_{>i-1}^\ddagger, f)\,d\theta.
\end{equation}
The normalized angular shapes at the next step are then obtained as
\begin{equation}
    P^L_{i+1}(\theta; \boldsymbol{l}_{<i+1}^\ddagger, f) = \frac{\hat{L}_{i+1}(\theta; \boldsymbol{l}_{<i+1}^\ddagger, f)}{\hat{A}_{i+1}^L(\boldsymbol{l}_{<i+1}^\ddagger, f)}, \quad
    P^R_{i-1}(\theta; \boldsymbol{l}_{>i-1}^\ddagger, f) = \frac{\hat{R}_{i-1}(\theta; \boldsymbol{l}_{>i-1}^\ddagger, f)}{\hat{A}_{i-1}^R(\boldsymbol{l}_{>i-1}^\ddagger, f)}.
\end{equation}
The same norms also update the accumulated magnitudes according to
\begin{eqnarray}
    A_{i+1}^L(\boldsymbol{l}_{<i+1}^\ddagger, f) &=& A_i^L(\boldsymbol{l}_{<i}^\ddagger, f)\hat{A}_{i+1}^L(\boldsymbol{l}_{<i+1}^\ddagger, f), \\ \nonumber
    A_{i-1}^R(\boldsymbol{l}_{>i-1}^\ddagger, f) &=& A_i^R(\boldsymbol{l}_{>i}^\ddagger, f) \hat{A}_{i-1}^R(\boldsymbol{l}_{>i-1}^\ddagger, f).
\end{eqnarray}
In practice, the accumulated factors $A_i^L$ and $A_i^R$ are stored in logarithmic form to avoid overflow or underflow, while $P_i^L$ and $P_i^R$ remain normalized angular distributions.

Substituting the factorized messages into Eq. (\ref{Seq:G_TM_original}) gives
\begin{equation}
    \label{Seq:G_TM_rescaled}
    \mathcal{G}_i(l; \boldsymbol{l}_{-i}^\ddagger, f) = 2\pi A_i^L(\boldsymbol{l}_{<i}^\ddagger, f) A_i^R(\boldsymbol{l}_{>i}^\ddagger, f) \int_0^\pi P^L_i(\theta; \boldsymbol{l}_{<i}^\ddagger, f) J(l, \theta; f) P^R_i(\theta; \boldsymbol{l}_{>i}^\ddagger, f)\,d\theta.
\end{equation}
The prefactor involving $A_i^L$ and $A_i^R$ is independent of $l$. The nontrivial $l$-dependence is contained in the fixed-length local factor $J(l,\theta;f)$ and its angular average. It is therefore convenient to define the angular weight
\begin{equation}
    w_i(\theta; \boldsymbol{l}_{-i}^\ddagger, f) = P^L_i(\theta; \boldsymbol{l}_{<i}^\ddagger, f) P^R_i(\theta; \boldsymbol{l}_{>i}^\ddagger, f),
\end{equation}
which encodes the bond-specific orientational statistics imposed by the adjacent subchains when the polar angle of bond $i$ is $\theta$.

At zero force, this angular weight has a particularly simple form. As noted in Sec. S1.1, the polar axis used to define $\theta$ is arbitrary in this limit. The force-dependent term in the configurational energy vanishes, leaving only bond-stretching terms, which depend on the bond lengths, and bending terms, which depend on the relative angles between neighboring bonds. The system is therefore invariant under global rotations. Since the rupture thresholds constrain only bond lengths, they do not break this symmetry. The adjacent subchains can impose correlations between neighboring bond directions, but they cannot favor any absolute polar angle of bond $i$. Hence the normalized message shapes are constant in $\theta$. Together with the normalization introduced above, this gives
\begin{equation}
    P^L_{i}(\theta; \boldsymbol{l}_{<i}^\ddagger, 0) = 
    P^R_{i}(\theta; \boldsymbol{l}_{>i}^\ddagger, 0) = \frac{1}{\pi}.
\end{equation}
Consequently,
\begin{equation}
    w_i(\theta; \boldsymbol{l}_{-i}^\ddagger, 0) = \frac{1}{\pi^2}.
\end{equation}
Thus, at zero force, the angular weight is independent of $\theta$ for every bond index. Bending correlations are still present, but rotational symmetry prevents them from producing a single-bond absolute polar-angle bias.

Using the definition of $J(l, \theta; f)$ in Eq. (\ref{Seq:J_l_theta}), the PMF, up to an $l$-independent additive constant, can be written as
\begin{equation}
\label{Seq:PMF_with_angular_weight}
    \mathcal{W}_i(l; \boldsymbol{l}_{-i}^\ddagger,f) = v_\mathrm{str}(l) - {\beta}^{-1} \ln{\left[l^2 \int_0^\pi w_i(\theta; \boldsymbol{l}_{-i}^\ddagger, f) e^{\beta fl\cos{\theta}} \sin{\theta}\,d\theta \right]}.
\end{equation}
The omitted additive constant does not affect the barrier location, barrier height, or the bond-resolved TST rate. This expression separates the bare stretching contribution from the orientational and entropic contributions encoded in the angular average.

The same rescaled messages also give the intact-basin configurational partition function. Substituting the factorized messages into Eq. (\ref{Seq:ZRconf_original}) yields
\begin{equation}
    \label{Seq:ZRconf_rescaled}
    Z_\mathcal{R}^\mathrm{conf}(\boldsymbol{l}^\ddagger; f) = 2\pi A_i^L(\boldsymbol{l}_{<i}^\ddagger, f) A_i^R(\boldsymbol{l}_{>i}^\ddagger, f) \int_0^\pi P^L_i(\theta; \boldsymbol{l}_{<i}^\ddagger, f) I_i(\theta; l_i^\ddagger, f) P^R_i(\theta; \boldsymbol{l}_{>i}^\ddagger, f)\,d\theta.
\end{equation}
The remaining angular integral is numerically well behaved, while the accumulated magnitudes are stored separately. The independence of $Z_\mathcal{R}^\mathrm{conf}(\boldsymbol{l}^\ddagger;f)$ from the chosen bond index $i$ provides a useful numerical check on the rescaled TM implementation.

For numerical evaluation in the variational TST scheme, one proceeds self-consistently as follows.
\begin{enumerate}
    \item Precompute the angular coupling kernel $Q(\theta,\theta')$ on the chosen angular grid.
    \item Choose an initial guess for the bond-dependent threshold set $\boldsymbol{l}^\ddagger = \{l_j^\ddagger\}$, for example a common initial value $l_\mathrm{guess}^\ddagger$.
    \item For the current threshold set, construct the local intact weights $I_j(\theta;l_j^\ddagger,f)$.
    \item Propagate the rescaled forward and backward messages, while computing the normalized angular shapes $P_i^L(\theta; \boldsymbol{l}_{<i}^\ddagger, f)$ and $P_i^R(\theta; \boldsymbol{l}_{>i}^\ddagger, f)$ and accumulating the logarithmic normalization factors $\ln A_i^L(\boldsymbol{l}_{<i}^\ddagger, f)$ and $\ln A_i^R(\boldsymbol{l}_{>i}^\ddagger, f)$.
    \item Assemble the angular weights $w_i(\theta; \boldsymbol{l}_{-i}^\ddagger, f)$ and construct the corresponding PMFs $\mathcal{W}_i(l; \boldsymbol{l}_{-i}^\ddagger, f)$.
    \item For each bond $i$, locate the PMF barrier position $l_i^\mathrm{b}(\boldsymbol{l}_{-i}^\ddagger, f)$ and set the corresponding threshold to $l_i^\ddagger = l_i^\mathrm{b}(\boldsymbol{l}_{-i}^\ddagger, f)$.
    \item Repeat steps 3–6 until the threshold set has converged.
\end{enumerate}
After convergence, the resulting thresholds are denoted by $l_{i,\mathrm{opt}}^\ddagger(f)$, and the corresponding converged PMFs are written as $\mathcal{W}_i(l;f)$. Substituting these optimized thresholds and PMFs into the bond-level TST expression, Eq. (\ref{Sqe:k_bond_with_W}), gives $k_{\mathrm{bond},i}^\mathrm{opt}(f)$, and the optimal chain-scission rate follows as $k_\mathrm{chain}^\mathrm{opt}(f) = \sum_{i=1}^N k_{\mathrm{bond},i}^\mathrm{opt}(f)$. The rescaled TM implementation therefore provides a numerically stable route to the self-consistent bond-dependent thresholds and the corresponding chain-scission rate for long chains.

\section{S3 Reduced reference models}
Before turning to the full numerical analysis, it is useful to consider two reduced models for which the PMF can be obtained without the self-consistent TM iteration required in the general bending-coupled chain. The first is the freely jointed (FJ) limit, which retains the three-dimensional bond-vector measure and orientational entropy but removes bending correlations. The second is the one-dimensional (1D) reference model, which removes all orientational degrees of freedom and isolates the purely energetic tensile tilt of the stretching potential. These two models provide internal consistency checks for the bond-resolved TST framework of Sec. S1 and the TM construction of Sec. S2.

\subsection{S3.1 Freely jointed limit}
We first consider the FJ limit, obtained by switching off the bending interaction, $v_\mathrm{ben}(\phi) = 0$. Neighboring bond orientations are then statistically independent, and the configurational energy reduces to a sum of uncoupled single-bond contributions,
\begin{equation}
    U_\mathrm{FJ}(\{l_i,\theta_i,\varphi_i\}_{i=1}^N;f) = \sum_{i=1}^{N} \left[v_\mathrm{str}{(l_i)} - fl_i \cos{\theta_i}\right].
\end{equation}
The configurational statistics therefore factorize \cite{manca2012elasticity,buche2022freely}. For a prescribed common threshold $l^\ddagger$, define the single-bond intact weight
\begin{equation}
    I_\mathrm{FJ}(l^\ddagger,f) = 2\pi \int_0^{l^\ddagger} e^{-\beta v_\mathrm{str}{(l)}} l^2 \left[\int_0^\pi e^{\beta fl \cos{\theta}}\sin{\theta}\,d\theta\right]\,dl,
\end{equation}
where the factor $2\pi$ comes from the free azimuthal integration. The constrained configurational weight at fixed $l_i = l$ is identical for all bonds:
\begin{equation}
    \mathcal{G}_\mathrm{FJ}(l; l^\ddagger, f) = 2\pi [I_\mathrm{FJ}(l^\ddagger, f)]^{N-1} e^{-\beta v_\mathrm{str}{(l)}} l^2 \left[\int_0^\pi e^{\beta fl \cos{\theta}}\sin{\theta}\,d\theta\right].
\end{equation}
Since $2\pi [I_\mathrm{FJ}(l^\ddagger, f)]^{N-1}$ is independent of $l$, it only shifts the PMF by a constant. Dropping this constant yields the common, threshold-independent FJ PMF,
\begin{equation}
\label{Seq:PMF_FJ}
    \mathcal{W}_\mathrm{FJ}(l;f) = v_\mathrm{str}(l) - {\beta}^{-1} \ln{\left[l^2 \int_0^\pi e^{\beta fl\cos{\theta}} \sin{\theta}\,d\theta \right]}. 
\end{equation}
The angular integral can be evaluated analytically, giving
\begin{equation}
    \mathcal{W}_\mathrm{FJ}(l;f) = v_\mathrm{str}(l) - {\beta}^{-1} \ln{\left[2l^2\frac{\sinh{(\beta fl)}}{\beta fl} \right]}.
\end{equation}
Compared with the general PMF expression in Eq. (\ref{Seq:PMF_with_angular_weight}), the FJ limit corresponds to a constant angular weight, because the surrounding subchains do not impose any additional $\theta$-dependent constraint on the selected bond. We set this constant to $w_\mathrm{FJ}(\theta) = 1$, since any constant prefactor only produces an $l$-independent shift in the PMF. This weight is therefore bond independent and independent of both rupture thresholds and applied force. Thus, the FJ PMF is determined only by the bare stretching potential, the radial bond-vector measure, and the force-biased orientational sampling of an isolated bond.

The FJ bonded minimum $l_\mathrm{FJ}^\mathrm{m}(f)$ and barrier top $l_\mathrm{FJ}^\mathrm{b}(f)$ are determined by the stationarity and curvature conditions on $\mathcal{W}_\mathrm{FJ}(l;f)$, as in Eq. (\ref{seq:stationarity_and_curvature_condition}). The associated barrier height is therefore $\Delta \mathcal W_\mathrm{FJ}(f) = \mathcal{W}_\mathrm{FJ}(l_\mathrm{FJ}^\mathrm{b}(f);f) - \mathcal{W}_\mathrm{FJ}(l_\mathrm{FJ}^\mathrm{m}(f);f)$. The PMF-based threshold construction then gives the optimized common rupture threshold $l_\mathrm{opt}^\ddagger(f) = l_\mathrm{FJ}^\mathrm{b}(f)$. Substitution into the bond-resolved TST expression gives
\begin{equation}
    \label{Seq:k_bond_with_W_FJ}
    k_{\mathrm{bond},i}^\mathrm{FJ,opt}(f) = \frac{1}{\sqrt{2\pi m_i^l \beta}} \frac{\displaystyle e^{-\beta \mathcal{W}_\mathrm{FJ}(l_\mathrm{FJ}^\mathrm{b}(f);f)}}{\displaystyle \int_0^{l_\mathrm{FJ}^\mathrm{b}(f)} e^{-\beta \mathcal{W}_\mathrm{FJ}(l;f)}\,dl}.
\end{equation}
Because the PMF contribution is identical for all bonds, the remaining bond dependence comes only from the kinetic prefactor. For bonds $i \geq 2$, $m_i^l = m/2$, and their common optimal rupture rate is denoted by $k_\mathrm{FJ}^\mathrm{opt}(f)$. For the first bond, $m_1^l = m$, so its larger effective mass reduces the kinetic prefactor, giving $k_{\mathrm{bond},1}^\mathrm{FJ,opt}(f) = k_\mathrm{FJ}^\mathrm{opt}(f)/{\sqrt{2}}$. Summing over all rupture channels gives
\begin{equation}
    k_\mathrm{chain}^\mathrm{FJ,opt}(f) = \sum_{i=1}^N k_{\mathrm{bond},i}^\mathrm{FJ,opt}(f) = \left(N - 1 + \frac{1}{\sqrt{2}}\right) k_\mathrm{FJ}^\mathrm{opt}(f) \approx N k_\mathrm{FJ}^\mathrm{opt}(f),
\end{equation}
where the approximation becomes accurate for large $N$. This limit shows that, once bending correlations are removed, the chain behaves as nearly $N$ equivalent rupture channels, apart from the fixed-end kinetic correction to the first bond. The resulting leading linear scaling with $N$ is consistent with previous results for breakable FJ chains \cite{buche2021chain}.

\subsection{S3.2 One-dimensional reference model}
As a reduced counterpart to the three-dimensional model, we also consider a 1D reference model (Fig. 1a of the main text). In this model, all orientational degrees of freedom are removed, and each bond is constrained to lie along the force direction, providing a reduced setting closely related to earlier transition-state treatments of force-loaded chains \cite{sebastian1999breaking,puthur2002theory}. The bond-resolved TST structure remains the same as in Sec. S1: the configurational contribution is described by a bond-length PMF, and the kinetic prefactor is determined by the associated velocity coordinate. The simplification is purely configurational, since the radial bond-vector measure, angular averaging, and TM propagation are all absent.

The 1D configurational energy is
\begin{equation}
    U_\mathrm{1D}(\{l_i\}_{i=1}^N; f) = \sum_{i=1}^N [v_\mathrm{str}(l_i) - fl_i].
\end{equation}
Because this energy is additive over bonds, the configurational statistics factorize. For a prescribed common threshold $l^\ddagger$, the single-bond intact weight is
\begin{equation}
    I_\mathrm{1D}(l^\ddagger, f) = \int_0^{l^\ddagger} e^{-\beta [v_\mathrm{str}(l) - fl]}\,dl.
\end{equation}
The constrained configurational weight at fixed $l_i = l$ is identical for all bonds:
\begin{equation}
    \mathcal{G}_\mathrm{1D}(l; l^\ddagger, f) = [I_\mathrm{1D}(l^\ddagger, f)]^{N-1} e^{-\beta [v_\mathrm{str}(l) - fl]}.
\end{equation}
The factor $[I_\mathrm{1D}(l^\ddagger, f)]^{N-1}$ is independent of $l$ and therefore only shifts the PMF by a constant. Dropping this constant yields the common, threshold-independent 1D PMF,
\begin{equation}
    \mathcal{W}_\mathrm{1D}(l;f) = v_\mathrm{str}(l) - fl.
\end{equation}
In contrast to the general 3D PMF in Eq. (\ref{Seq:PMF_with_angular_weight}), the 1D PMF contains neither the radial Jacobian contribution nor any force-dependent angular average. It therefore isolates the purely energetic effect of tilting the stretching potential by the applied force.

The 1D bonded minimum $l_\mathrm{1D}^\mathrm{m}(f)$ and barrier top $l_\mathrm{1D}^\mathrm{b}(f)$ are determined by the stationarity and curvature conditions on $\mathcal{W}_\mathrm{1D}(l;f)$, as in Eq. (\ref{seq:stationarity_and_curvature_condition}). For the Morse stretching potential, $v_\mathrm{str}(l) = D_e[1 - e^{-a(l-l_e)}]^2$, these stationary points can be obtained analytically \cite{charan2021aging}. For $0 < f < aD_e/2$, they are
\begin{equation}
    l_\mathrm{1D}^\mathrm{m}(f) = l_e - \frac{1}{a} \ln{\left[\frac{1 + \sqrt{1 - 2f/(aD_e)}}{2}\right]}, \quad
    l_\mathrm{1D}^\mathrm{b}(f) = l_e - \frac{1}{a} \ln{\left[\frac{1 - \sqrt{1 - 2f/(aD_e)}}{2}\right]}.
\end{equation}
At zero force, the PMF reduces to the bare Morse potential, $\mathcal{W}_\mathrm{1D}(l;0) = v_\mathrm{str}(l)$. The bonded minimum remains at
$l_\mathrm{1D}^\mathrm{m}(0)=l_e$, but no finite stationary barrier top exists. Instead, $l_\mathrm{1D}^\mathrm{b}(f)\rightarrow\infty$ as $f\rightarrow0^+$, and the PMF approaches the dissociation plateau $D_e$ asymptotically.

Within this finite-force regime, the PMF-based threshold construction gives the optimized common rupture threshold $l^\ddagger_\mathrm{opt}(f) = l_\mathrm{1D}^\mathrm{b}(f)$. Substitution into the bond-resolved TST expression yields
\begin{equation}
    \label{seq:TST_rate_full_1D}
    k_{\mathrm{bond},i}^\mathrm{1D,opt}(f) = \frac{1}{\sqrt{2\pi m_i^l \beta}} \frac{\displaystyle e^{-\beta \mathcal{W}_\mathrm{1D}(l_\mathrm{1D}^\mathrm{b}(f);f)}}{\displaystyle \int_0^{l_\mathrm{1D}^\mathrm{b}(f)} e^{-\beta \mathcal{W}_\mathrm{1D}(l;f)}\,dl}.
\end{equation}
The kinetic prefactor retains the same structure as in Sec. S1 because the 1D reduction removes only configurational degrees of freedom and leaves the bond-length velocity coordinate unchanged. Accordingly, $m_1^l = m$ for the first bond, whereas $m_i^l = m/2$ for bonds $i \geq 2$.

The corresponding activation barrier is $\Delta \mathcal{W}_\mathrm{1D}(f) = \mathcal{W}_\mathrm{1D}(l_\mathrm{1D}^\mathrm{b}(f);f) - \mathcal{W}_\mathrm{1D}(l_\mathrm{1D}^\mathrm{m}(f);f)$, which gives the analytic result for a linearly tilted Morse potential \cite{khodayeki2022force}
\begin{equation}
    \Delta \mathcal{W}_\mathrm{1D}(f) = D_e \sqrt{1 - \frac{2f}{aD_e}} + \frac{f}{a} \ln{\left(\frac{1-\sqrt{1 - 2f/(aD_e)}}{1+\sqrt{1 - 2f/(aD_e)}}\right)}.
\end{equation}
It approaches $D_e$ as $f \rightarrow 0^+$ and decreases monotonically with increasing force. It vanishes at the 1D critical force, $f_c^\mathrm{1D} = aD_e/2$, where the bonded minimum and barrier top coincide. In the high-barrier regime, $\beta \Delta \mathcal{W}_\mathrm{1D}(f) \gg 1$, a harmonic approximation about the bonded minimum gives the Arrhenius form
\begin{equation}
    \label{seq:Arrhenius_form_1D}
    k_{\mathrm{bond},i}^{\mathrm{1D,opt}}(f) \approx \frac{\omega_{\mathrm{1D},i}^\mathrm{m}(f)}{2\pi} e^{-\beta \Delta \mathcal{W}_\mathrm{1D}(f)}, 
\end{equation}
where 
\begin{equation}
    m_i^l[\omega_{\mathrm{1D},i}^\mathrm{m}(f)]^2 = \left.\frac{\partial^2 \mathcal{W}_\mathrm{1D}(l;f)}{\partial l^2}\right\vert_{l = l_\mathrm{1D}^\mathrm{m}(f)} 
    = a^2 D_e \left(1 + \sqrt{1 - \frac{2f}{aD_e}}\right) \sqrt{1 - \frac{2f}{aD_e}}.
\end{equation}

Because no finite optimized threshold exists at $f = 0$, we define the zero-force 1D bond-resolved rate by the $f \rightarrow 0^+$ limit of the Arrhenius form:
\begin{equation}
    \label{seq:zero_force_rate_1D}
    k_{\mathrm{bond},i}^{\mathrm{1D,opt}}(0) \equiv \lim_{f \rightarrow 0^+} \frac{\omega_{\mathrm{1D},i}^\mathrm{m}(f)}{2\pi} e^{-\beta \Delta \mathcal{W}_\mathrm{1D}(f)} = \frac{a}{2\pi} \sqrt{\frac{2 D_e}{m_i^l}} e^{-\beta D_e}.
\end{equation}

Let $k_\mathrm{1D}^\mathrm{opt}(f)$ denote the common optimal rate for bonds $i \geq 2$. The larger effective mass of the first bond reduces its kinetic prefactor, giving $k_{\mathrm{bond},1}^\mathrm{1D,opt}(f) = k_\mathrm{1D}^\mathrm{opt}(f)/{\sqrt{2}}$. Summing over all rupture channels then gives
\begin{equation}
    k_\mathrm{chain}^\mathrm{1D,opt}(f) = \sum_{i=1}^N k_{\mathrm{bond},i}^\mathrm{1D,opt}(f) = \left(N - 1 + \frac{1}{\sqrt{2}}\right) k_\mathrm{1D}^\mathrm{opt}(f) \approx N k_\mathrm{1D}^\mathrm{opt}(f),
\end{equation}
where the approximation becomes accurate for large $N$. Thus, the 1D reference retains the bond-length kinetic structure while removing the radial measure and orientational contributions to the PMF.

\section{S4 Numerical parameterization and reduced units}
The numerical calculations in the main text and in Secs. S5--S7 are parameterized in reduced units. The local bonded interactions are characterized by the reduced stretching parameters $\beta D_e$ and $a l_e$, the equilibrium angle $\phi_e$, and the reduced bending stiffness $\beta k_\phi \pi^2$, while the applied force is reported as $fl_e/D_e$.

The Boltzmann constant is $k_B = 1.38\times10^{-23}\mathrm{J/K}$, and the temperature is taken as $T = 293.15\mathrm{K}$. This gives
\begin{equation}
    k_B T \approx 4.05 \times 10^{-21} \mathrm{J}, \quad 
    \beta = (k_B T)^{-1} \approx 2.47 \times 10^{20} \mathrm{J}^{-1}.
\end{equation}

We use representative carbon–carbon bonded-interaction parameters. For the stretching interaction, the Morse parameters are $l_e = 1.525\mathrm{\AA}$, $D_e = 1.13\times10^{-18}\mathrm{J}$, and $a = 1.409\mathrm{\AA}^{-1}$, taken from Ref. \cite{shannon2019anharmonic}. These values correspond to
\begin{equation}
    \beta D_e \approx 279, \quad a l_e \approx 2.15.
\end{equation}
For the bending interaction, the equilibrium bond angle is set to $\phi_e=69^{\circ}$, and the representative bending stiffness is $k_\phi=7.47\times10^{-19}\mathrm{J}/\mathrm{rad}^2$, taken from Ref. \cite{sorensen1988prediction}. This stiffness corresponds to
\begin{equation}
    \beta k_\phi \pi^2 \approx 1.82 \times 10^3.
\end{equation}

The absolute scale of the TST rates is set by the kinetic prefactor $1/\sqrt{2\pi m_i^l\beta}$. We take the atomic mass of carbon to be $m = 1.99\times10^{-26} \mathrm{kg}$. For the fixed-end bond, the effective bond-length mass is $m_1^l = m$, whereas for the remaining bonds we use the reduced mass of two equal carbon atoms, $m_i^l = m/2$ for $i \geq 2$. The corresponding velocity prefactors are
\begin{equation}
    1.80 \times 10^{2}\mathrm{m/s}~~\mathrm{for}~~i = 1, \quad
    2.54 \times 10^{2} \mathrm{m/s}~~\mathrm{for}~~i \geq 2.
\end{equation}
When the rate expression is evaluated using the reduced bond length $l_i/l_e$, these velocity prefactors are divided by $l_e$, giving frequency prefactors
\begin{equation}
    1.18 \times 10^{12}\mathrm{s}^{-1}~~\mathrm{for}~~i = 1, \quad
    1.67 \times 10^{12} \mathrm{s}^{-1}~~\mathrm{for}~~i \geq 2.
\end{equation}

Unless otherwise stated, this parameter set and the corresponding carbon-mass kinetic prefactors are used in the numerical calculations. In parameter-sensitivity tests, individual quantities such as the bending stiffness are varied as indicated.

\section{S5 Angular-weight profiles}
To clarify the conformational origin of the bond-resolved PMFs, we examine the angular weights entering Eq. (\ref{Seq:PMF_with_angular_weight}). Throughout this section, the angular weights are the self-consistently converged weights evaluated using the optimized rupture-threshold set, $\boldsymbol{l}_\mathrm{opt}^\ddagger(f)$. For compactness, we write $w_i(\theta; \boldsymbol{l}_\mathrm{opt}^\ddagger(f), f)$ as $w_i(\theta; f)$. The angular weight isolates the orientational constraint imposed by the adjacent subchains and is not itself the full polar-angle probability density of bond $i$, which additionally contains the local force-bias factor $e^{\beta f l \cos{\theta}}$ and the geometrical measure $\sin{\theta}$.

To compare angular-weight profiles across bond positions, bending stiffnesses, and applied forces, we introduce the mean-normalized angular weight
\begin{equation}
\label{Seq:mean-normalization}
    \tilde{w}_i(\theta; f) = \frac{\pi w_i(\theta; f)}{\displaystyle \int_0^\pi w_i(\theta; f)\,d\theta}.
\end{equation}
This normalization sets the mean value of $\tilde{w}_i$ over $\theta \in [0, \pi]$ to unity, thereby isolating the shape of its angular dependence. For fixed bond index $i$ and applied force $f$, the normalization factor is independent of the local bond length $l$. It therefore contributes only an $l$-independent shift to the PMF and does not affect the barrier location, barrier height, or corresponding bond-resolved rate. The normalization also preserves the FJ limit as the constant reference $\tilde{w}_\mathrm{FJ}(\theta; f) = 1$.

\subsection{S5.1 Spatial variation}
We first examine how the angular-weight profiles vary with bond position along the chain.  The calculation uses $N = 100$, the carbon-carbon-based parameter set of Sec. S4, and $f l_e/D_e = 0.2$. 

\begin{figure} 
    \includegraphics{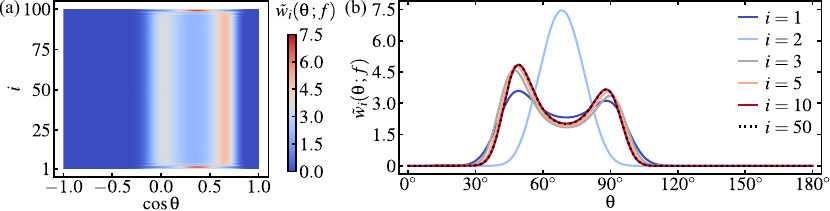}
    \caption{Bond-position dependence of angular weights. (a) Color map of the mean-normalized angular weight $\tilde{w}_i(\theta; f)$ as a function of bond index $i$ and bond orientation $\cos{\theta}$ at $fl_e/D_e = 0.2$. (b) Representative profiles of $\tilde{w}_i(\theta; f)$ as functions of $\theta$ for selected bond indices $i = 1$, $2$, $3$, $5$, $10$, and $50$. Parameters: $N=100$, $al_e=2.15$, $\beta D_e=279$, $\phi_e = 69^\circ$, and $\beta k_\phi\pi^2 = 1.82 \times 10^3$.}
    \label{sfig:DBA_angular_weight_representative}
\end{figure}

Fig. \ref{sfig:DBA_angular_weight_representative}a shows a color map of $\tilde{w}_i(\theta; f)$ as a function of bond index $i$ and bond orientation $\cos{\theta}$. The color map reveals a clear separation between near-end boundary layers and a bulk-like interior region. Because the configurational interactions are homogeneous and the bending correlations are truncated identically at the two chain ends, the angular-weight profiles are mirror-symmetric about the chain midpoint, $\tilde{w}_i(\theta; f) = \tilde{w}_{N+1-i}(\theta; f)$. It is therefore sufficient to examine the evolution of $\tilde{w}_i(\theta; f)$ from one terminus toward the midpoint. The pronounced variation near each terminus originates from the truncation of bending correlations at the chain boundary. For the terminal bond, orientational constraints are transmitted from only one adjacent subchain. For nearby nonterminal bonds, constraints are transmitted from both sides, but the end-facing and interior-facing subchains have unequal lengths and configurational environments. As this boundary-induced constraint asymmetry decays with distance from the terminus, $\tilde{w}_i(\theta; f)$ approaches a nearly position-independent bulk-like interior profile, which we denote by $\tilde{w}_\infty(\theta; f)$. 

Fig. \ref{sfig:DBA_angular_weight_representative}b compares representative profiles for $i=1$, $2$, $3$, $5$, $10$, and $50$, tracing the relaxation of the boundary-induced angular-weight perturbation from one terminus toward the chain midpoint. The terminal bond $i = 1$ exhibits a double-peaked profile, with maxima near $\theta \approx 49^\circ$ and $\theta \approx 89^\circ$. These peaks represent two absolute bond orientations compatible with the preferred zigzag-like relative-angle geometry of the adjacent subchain. The smaller-$\theta$ peak is slightly higher than the larger-$\theta$ peak because the applied force biases the surrounding subchain conformations toward the pulling direction, indirectly enhancing the corresponding orientational branch in the angular weight. Orientations outside these two branches, especially for $\theta \lesssim 20^\circ$ and $\theta \gtrsim 120^\circ$, are strongly suppressed by bending compatibility with the adjacent subchain.

Moving one bond inward, the first nonterminal bond $i = 2$ exhibits a single dominant peak near $\theta \approx 69^\circ$, numerically close to the equilibrium bond angle $\phi_e$. Although this bond is coupled to neighboring bonds on both sides, its adjacent subchains are highly unequal: the end-facing side contains only the terminal bond, whereas the interior-facing side contains the remainder of the chain. The combination of two-sided bending compatibility and this strong boundary-induced constraint asymmetry produces a comparatively localized, single-peaked angular-weight profile.

Further into the chain, the profiles for $i=3$, $5$, $10$, and $50$ regain a double-peaked structure associated with zigzag-compatible orientations and progressively approach the bulk-like interior profile $\tilde{w}_\infty(\theta; f)$. Along this sequence, the two peaks become slightly sharper and move slightly closer to each other. Finally, the angular-weight profiles converge toward their bulk-like forms, while the smaller-$\theta$ branch becomes increasingly prominent. This evolution reflects the decay of the boundary-induced constraint asymmetry and the increasingly apparent bias of the surrounding subchain conformations toward the pulling direction under force.

Taken together, the sequence from $i = 1$ to $i = 50$ shows that $\tilde{w}_i(\theta; f)$ relaxes toward $\tilde{w}_\infty(\theta; f)$ through a nonmonotonic reorganization localized near the chain ends. This reorganization reflects the interplay among bending compatibility, boundary-induced constraint asymmetry, and force-biased conformations of the surrounding subchains.

\subsection{S5.2 Dependence on bending stiffness}
We next examine how the angular-weight profiles change with bending stiffness, while keeping all remaining parameters fixed at the values used in Sec. S5.1. At $k_\phi = 0$, the model reduces to the freely jointed limit discussed in Sec. S3.1, for which neighboring bond orientations are statistically independent and the mean-normalized angular weight is constant, $\tilde{w}_\mathrm{FJ}(\theta; f) = 1$. Deviations of $\tilde{w}_i(\theta; f)$ from unity at finite $k_\phi$ therefore reflect the orientational constraints generated by bending correlations.

\begin{figure} 
    \includegraphics{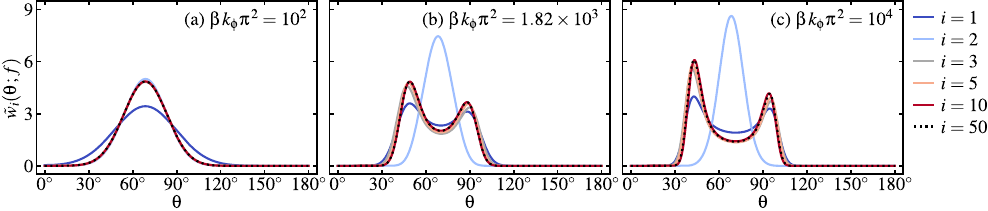}
    \caption{Bending-stiffness dependence of angular weights. Representative profiles of $\tilde{w}_i(\theta; f)$ at $f l_e/D_e = 0.2$ for selected bond indices $i = 1$, $2$, $3$, $5$, $10$, and $50$, shown for bending stiffnesses (a) $\beta k_\phi\pi^2 = 10^2$, (b) $\beta k_\phi\pi^2 = 1.82 \times 10^3$, and (c) $\beta k_\phi\pi^2 = 10^4$. Other parameters: $N = 100$, $a l_e = 2.15$, $\beta D_e = 279$, and $\phi_e = 69^\circ$.}
    \label{sfig:DBA_angular_weight_vary_kb}
\end{figure}

Fig. \ref{sfig:DBA_angular_weight_vary_kb} compares representative profiles of $\tilde{w}_i(\theta; f)$ for selected bond indices $i = 1$, $2$, $3$, $5$, $10$, and $50$ at three bending stiffnesses, $\beta k_\phi\pi^2 = 10^2$, $1.82 \times 10^3$, and $10^4$. For weak bending, $\beta k_\phi\pi^2 = 10^2$, the angular-weight profiles are single-peaked for all selected bond positions, with maxima near $\theta \approx 69^\circ$ (Fig. \ref{sfig:DBA_angular_weight_vary_kb}a). The finite bending stiffness produces a clear preferred angular range, but the correlations remain too weak to resolve distinct zigzag-compatible branches. The terminal-bond profile is lower and broader than those of the nonterminal bonds because orientational constraints are transmitted from only one adjacent subchain. By contrast, the profiles for $i = 2$, $3$, $5$, $10$, and $50$ nearly overlap, indicating that the boundary-induced constraint asymmetry decays rapidly beyond the terminal bond in the weak-bending regime.

At intermediate stiffness, $\beta k_\phi\pi^2 = 1.82 \times 10^3$, the spatial pattern analyzed in Sec. S5.1 becomes clearly resolved (Fig. \ref{sfig:DBA_angular_weight_vary_kb}b). The terminal bond exhibits a double-peaked profile with maxima near $\theta \approx 49^\circ$ and $\theta \approx 89^\circ$, whereas the first nonterminal bond retains a strongly localized single peak near $\theta \approx 69^\circ$. Bonds farther inward regain a double-peaked structure and converge toward the bulk-like interior profile, in which the smaller-$\theta$ branch is more prominent than the larger-$\theta$ branch. Compared with the weak-bending case, bending correlations are sufficiently strong to resolve both the boundary-induced spatial variation and the zigzag-compatible orientational branches.

Upon further increasing the stiffness to $\beta k_\phi\pi^2 = 10^4$, the same qualitative spatial pattern persists, but all selected angular-weight profiles become more sharply localized (Fig. \ref{sfig:DBA_angular_weight_vary_kb}c). For the double-peaked profiles, the two branches also become more widely separated, with maxima near $\theta \approx 43^\circ$ and $\theta \approx 94^\circ$. The sharper peaks indicate stronger angular localization, while the increased separation makes the two zigzag-compatible branches more distinctly resolved. The enhanced contrast between the near-end profiles and the bulk-like interior profile further demonstrates that bending stiffness amplifies the spatial variation of the angular weights.

Overall, increasing bending stiffness sharpens the angular-weight profiles and resolves distinct zigzag-compatible branches at most bond positions, while the first nonterminal bond retains a strongly localized single-peaked profile. The increasing stiffness also amplifies the contrast between the near-end region and the bulk-like interior.

\subsection{S5.3 Dependence on applied force}
We then examine how the angular-weight profiles change with applied force, while keeping all remaining parameters fixed at the values used in Sec. S5.1. As a reference, we first recall the zero-force result discussed in Sec. S2.2. At $f = 0$, global rotational symmetry prevents the adjacent subchains from favoring any absolute polar angle of a selected bond. The angular weight is therefore independent of $\theta$ for every bond index, and the mean normalization in Eq. (\ref{Seq:mean-normalization}) gives $\tilde{w}_i(\theta; 0) = 1$. Finite applied force breaks this rotational symmetry by biasing the surrounding subchain conformations toward the pulling direction. Through bending correlations, this subchain-level bias indirectly generates the nonuniform angular-weight profiles observed below.

\begin{figure} 
    \includegraphics{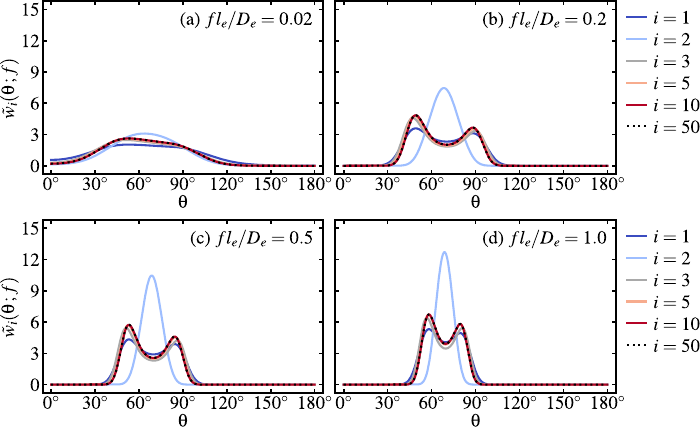}
    \caption{Force dependence of angular weights. Representative profiles of $\tilde{w}_i(\theta; f)$ for selected bond indices $i = 1$, $2$, $3$, $5$, $10$, and $50$, shown for applied forces (a) $f l_e/D_e = 0.02$, (b) $f l_e/D_e = 0.2$, (c) $f l_e/D_e = 0.5$, and (d) $f l_e/D_e = 1.0$. Other parameters: $N = 100$, $a l_e = 2.15$, $\beta D_e = 279$, $\phi_e = 69^\circ$, and $\beta k_\phi\pi^2 = 1.82 \times 10^3$.}
    \label{sfig:DBA_angular_weight_vary_f}
\end{figure}

Fig. \ref{sfig:DBA_angular_weight_vary_f} compares representative profiles of $\tilde{w}_i(\theta; f)$ for selected bond indices $i = 1$, $2$, $3$, $5$, $10$, and $50$ at four applied forces, $f l_e/D_e = 0.02$, $0.2$, $0.5$, and $1.0$. At the lowest force, $f l_e/D_e = 0.02$, the profiles deviate only slightly from the flat zero-force reference (Fig. \ref{sfig:DBA_angular_weight_vary_f}a). The first nonterminal bond $i = 2$ shows a broad maximum near $\theta \approx 65^\circ$, whereas the remaining profiles vary more gradually with $\theta$ and remain closely clustered throughout $\theta \in (30^\circ, 100^\circ)$. Appreciable depletion occurs only at large angles, particularly for $\theta \gtrsim 150^\circ$. These features reflect the weak bias produced by the applied force in the surrounding subchain conformations, and no distinct zigzag-compatible branches are resolved in this low-force regime.

At $f l_e/D_e = 0.2$, the spatial pattern analyzed in Sec. S5.1 becomes clearly resolved (Fig. \ref{sfig:DBA_angular_weight_vary_f}b). The first nonterminal bond develops a pronounced maximum near $\theta \approx 69^\circ$, while the terminal and interior profiles develop distinct double-peaked structures with maxima near $\theta \approx 49^\circ$ and $\theta \approx 89^\circ$. Orientations with $\theta \lesssim 20^\circ$ and $\theta \gtrsim 120^\circ$ are now strongly suppressed. Compared with the lowest-force case, the subchain-level force bias is now sufficient for both the zigzag-compatible branches and the unequal bending constraints at different bond positions to become apparent.

As the force is increased further to $f l_e/D_e = 0.5$ and $1.0$, the same qualitative spatial pattern persists, but all selected angular-weight profiles become more sharply localized (Figs. \ref{sfig:DBA_angular_weight_vary_f}c and \ref{sfig:DBA_angular_weight_vary_f}d). The first nonterminal bond retains a single maximum near $\theta \approx 69^\circ$, which becomes progressively narrower and higher with increasing force, whereas the terminal and interior profiles retain double-peaked structures whose branches sharpen and move toward one another. For the bulk-like interior profile, the maxima shift to $\theta \approx 53^\circ$ and $\theta \approx 85^\circ$ at $f l_e/D_e = 0.5$, and to $\theta \approx 58^\circ$ and $\theta \approx 80^\circ$ at $f l_e/D_e = 1.0$. Meanwhile, the depleted angular regions broaden to $\theta \lesssim 25^\circ$ and $\theta \gtrsim 115^\circ$ at $f l_e/D_e = 0.5$, and to $\theta \lesssim 30^\circ$ and $\theta \gtrsim 110^\circ$ at $f l_e/D_e = 1.0$. These changes reflect the increasing tendency of the surrounding subchain conformations to align with the pulling direction, which, through bending compatibility, progressively confines the angular weights to narrower intermediate-orientation regions.

Overall, increasing force makes the bending-mediated bond-position dependence increasingly apparent, progressively sharpens the angular-weight profiles, and narrows the favored intermediate-orientation ranges.

\section{S6 Bond-resolved activation barriers}
Using the mean-normalized angular weights introduced in Sec. S5, the converged bond-length PMF can be written, up to an $l$-independent additive constant, as
\begin{equation}
    \label{Seq:converged_PMF_with_angular_weight}
    \mathcal{W}_i(l; f) = v_\mathrm{str}(l) - {\beta}^{-1} \ln{\left[l^2 \int_0^\pi \tilde{w}_i(\theta; f) e^{\beta fl\cos{\theta}} \sin{\theta}\,d\theta \right]}.
\end{equation}
The bond-position dependence of $\tilde{w}_i(\theta; f)$ is transmitted through the angular integral to the PMFs and the corresponding activation barriers $\Delta \mathcal{W}_i(f)$. In this section, we examine these barrier profiles and their dependence on chain length, bending stiffness, and applied force.

\subsection{S6.1 Length-dependent convergence and boundary-layer formation}
Fig. 2b of the main text shows that, at finite bending stiffness and finite force, the bond-resolved activation barriers $\Delta \mathcal{W}_i(f)$ are mirror-symmetric about the chain midpoint, $\Delta \mathcal{W}_i(f) = \Delta \mathcal{W}_{N+1-i}(f)$. The terminal bonds have the lowest barriers, the first nonterminal bonds have the highest barriers, and bonds farther into the chain approach an interior plateau, denoted by $\Delta \mathcal{W}_\infty(f)$, through damped oscillations. This spatial structure originates from the bond-position dependence of the angular-weight profiles discussed in Sec. S5.1, which are likewise mirror-symmetric about the chain midpoint and undergo a nonmonotonic near-end reorganization before approaching a bulk-like interior form. We now examine how this long-chain barrier structure emerges as the chain length $N$ increases and identify the spatial range over which the chain ends perturb the barrier landscape. The calculations in this subsection use the carbon–carbon-based parameter set of Sec. S4 and a reduced force $f l_e/D_e = 0.2$.

\begin{figure} 
    \includegraphics{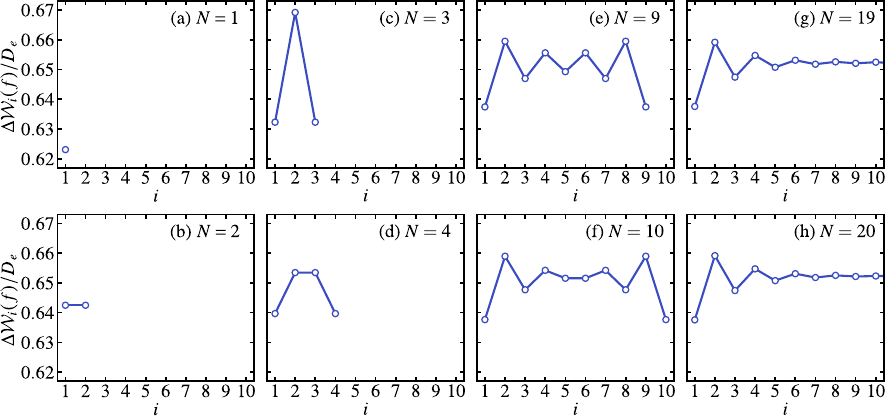}
    \caption{Bond-resolved activation barriers $\Delta \mathcal{W}_i(f)$ at $fl_e/D_e = 0.2$ for chains of varying length $N$. Panels (a-h) show $N = 1$, $2$, $3$, $4$, $9$, $10$, $19$, and $20$, respectively. Only bonds within the first ten positions from one chain end are shown; for $N \leq 10$, this corresponds to the entire chain. Parameters: $a l_e = 2.15$, $\beta D_e = 279$, $\phi_e = 69^\circ$, and $\beta k_\phi\pi^2 = 1.82 \times 10^3$.}
    \label{sfig:DBA_barrier_vary_N_small}
\end{figure}

Fig. \ref{sfig:DBA_barrier_vary_N_small} shows $\Delta \mathcal{W}_i(f)$ for representative short and intermediate chain lengths, with panels (a--h) corresponding to $N = 1$, $2$, $3$, $4$, $9$, $10$, $19$, and $20$, respectively. For $N = 1$, no bond angle exists and the bending energy is absent, so the activation barrier coincides with the freely jointed single-bond result. Bending correlations first enter at $N = 2$, where the single bond angle couples the two bond orientations. However, the two bonds remain equivalent and therefore have the same barrier. For $N = 3$ and $4$, the central bond or central bond pair has a higher barrier than the terminal bonds. The difference between the odd- and even-$N$ profiles suggests a short-chain parity effect associated with the accommodation of the preferred zigzag-like geometry between the two chain ends. By $N = 9$ and $10$, the profile already resembles the long-chain form: the barriers rise from their terminal values and approach an emerging interior plateau through damped oscillations. The profiles for $N = 19$ and $20$ are nearly indistinguishable over the displayed near-end range, indicating that the near-end barrier structure has already converged closely to its long-chain form. 

To compare chains of different lengths on a common spatial coordinate, we label each bond by its distance from the nearest chain end, $s \equiv \min(i, N+1-i)$, and denote the corresponding finite-chain barrier by $\Delta \mathcal{W}_s(f; N)$. Here, $s = 1$ labels the two terminal bonds, and increasing $s$ moves toward the chain interior. Mirror symmetry makes the two chain ends equivalent, so bonds at equal distance from opposite ends share the same value of $s$. For a given $s$, $\Delta \mathcal{W}_s(f; N)$ is defined only when $N \geq 2s -1$.

\begin{figure} 
    \includegraphics{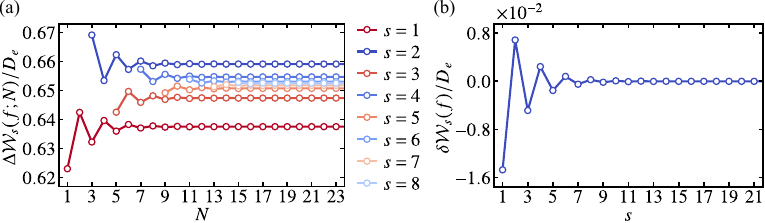}
    \caption{Convergence of near-end activation barriers and boundary deviations. (a) Chain-length dependence of the activation barriers $\Delta \mathcal{W}_s(f;N)$ at $fl_e/D_e = 0.2$, grouped by the distance from the nearest chain end, $s \equiv \min{(i, N + 1 -i)}$. (b) Converged near-end deviation $\delta \mathcal{W}_s(f) = \Delta \mathcal{W}_s(f) - \Delta \mathcal{W}_\infty(f)$ at the same force, evaluated using a representative long chain with $N = 100$. Parameters: $a l_e = 2.15$, $\beta D_e = 279$, $\phi_e = 69^\circ$, and $\beta k_\phi\pi^2 = 1.82 \times 10^3$.}
    \label{sfig:DBA_barrier_boundary_layer_convergence}
\end{figure}

Fig. \ref{sfig:DBA_barrier_boundary_layer_convergence}a plots $\Delta \mathcal{W}_s(f; N)$ as a function of $N$ for several fixed values of $s$. The condition $N \geq 2s -1$ sets the starting point of each curve. For short chains, the perturbations generated by the two ends overlap, and the barrier at fixed $s$ depends appreciably on the total chain length. As $N$ increases, the opposite end moves farther away and the local environment at fixed $s$ becomes insensitive to it. Consequently, $\Delta \mathcal{W}_s(f; N)$ approaches an $N$-independent limiting value. The chain length required for this convergence increases with $s$, because bonds farther from the terminus require a longer chain before their local environment becomes independent of the opposite end. We denote the converged value at fixed $s$ by $\Delta \mathcal{W}_s(f)$. Far from either end, this converged near-end profile approaches the interior plateau $\Delta \mathcal{W}_\infty(f)$.

To quantify the end-induced perturbation relative to the interior plateau, we define the converged boundary deviation 
\begin{equation}
    \delta \mathcal{W}_s(f) = \Delta \mathcal{W}_s(f) - \Delta \mathcal{W}_\infty(f).
\end{equation}
As shown in Fig. \ref{sfig:DBA_barrier_boundary_layer_convergence}b, $\delta \mathcal{W}_s(f)$ begins at a negative terminal value and approaches zero through damped oscillations as $s$ increases. This behavior reflects the progressive loss of end effects and the approach to the interior plateau. 

The magnitude of the terminal offset provides a natural measure of the amplitude of the end-induced perturbation. We therefore define the terminal barrier reduction as
\begin{equation}
    -\delta \mathcal{W}_1(f) \equiv \Delta \mathcal{W}_\infty(f) - \Delta \mathcal{W}_1(f).
\end{equation}
A larger value of $-\delta \mathcal{W}_1(f)$ indicates a stronger terminal-to-interior barrier contrast and therefore a greater kinetic preference for rupture near the chain ends.

The spatial extent of the end-induced perturbation is quantified by the boundary-layer length $\ell_b(f)$, defined as the largest end distance at which the normalized barrier deviation exceeds a prescribed tolerance:
\begin{equation}
    \ell_b(f) = \max\{s:|\delta \mathcal{W}_s(f)|/D_e > \varepsilon\}.
\end{equation}
We set $\ell_b(f) = 0$ if no bond exceeds the tolerance. Equivalently, all bonds satisfying $s > \ell_b(f)$ obey $|\delta \mathcal{W}_s(f)|/D_e \leq \varepsilon$. In the numerical analysis, we use $\varepsilon = 10^{-5}$. For the parameter set and force used in this subsection, the converged profile gives $\Delta \mathcal{W}_\infty(f) \approx 0.65 D_e$, $-\delta \mathcal{W}_1(f) \approx 0.015 D_e$, and $\ell_b(f) = 13$.

These three quantities provide a compact characterization of the long-chain barrier profile. The interior plateau $\Delta \mathcal{W}_\infty(f)$ sets the bulk-like activation barrier, the terminal reduction $-\delta \mathcal{W}_1(f)$ measures the amplitude of the end-induced perturbation, and $\ell_b(f)$ specifies its spatial extent. For a finite chain, the two terminal boundary layers cease to overlap once the chain is sufficiently long. When $N > 2\ell_b(f)$, an interior region satisfying $s > \ell_b(f)$ is present between them. Further increases in $N$ mainly add bonds to the interior plateau without altering the near-end structures. In this long-chain regime, the barrier profile consists of two terminal boundary layers separated by a bulk-like interior plateau.

\subsection{S6.2 Dependence on bending stiffness}
\begin{figure} 
    \includegraphics{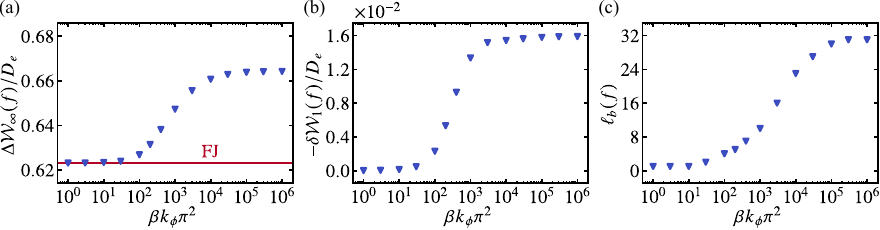}
    \caption{Extracted measures of the long-chain barrier profile as functions of bending stiffness. (a) Interior plateau barrier $\Delta \mathcal{W}_\infty(f)$. (b) Terminal barrier reduction $-\delta \mathcal{W}_1(f)$. (c) Boundary-layer length $\ell_b(f)$. The FJ limit is indicated by a horizontal reference line in panel (a); in panels (b) and (c), the corresponding values are $-\delta \mathcal{W}_1(f) = 0$ and $\ell_b(f) = 0$, respectively. All panels are evaluated at $fl_e/D_e = 0.2$. Other parameters: $a l_e = 2.15$, $\beta D_e = 279$, and $\phi_e = 69^\circ$.}
    \label{sfig:DBA_barrier_measures_vary_kb}
\end{figure}

We next examine how the bending stiffness modifies the long-chain barrier profile, while keeping all remaining parameters fixed at the values used in Sec. S6.1. Fig. \ref{sfig:DBA_barrier_measures_vary_kb} shows the stiffness dependence of the interior plateau barrier $\Delta \mathcal{W}_\infty(f)$, the terminal barrier reduction $-\delta \mathcal{W}_1(f)$, and the boundary-layer length $\ell_b(f)$ over the range $10^0 \leq \beta k_\phi\pi^2 \leq 10^6$.

Fig. \ref{sfig:DBA_barrier_measures_vary_kb}a characterizes the bulk-like barrier level through the interior plateau barrier $\Delta \mathcal{W}_\infty(f)$. At sufficiently low bending stiffness, $\beta k_\phi\pi^2 \lesssim 10^1$, $\Delta \mathcal{W}_\infty(f)$ approaches the FJ value, $\Delta \mathcal{W}_\mathrm{FJ}(f) \approx 0.62 D_e$. As the bending stiffness increases, $\Delta \mathcal{W}_\infty(f)$ rises and, for $\beta k_\phi\pi^2 \gtrsim 10^5$, approaches the fixed-bond-angle limiting value of approximately $0.66 D_e$. This trend originates from changes in the bulk-like angular-weight profile $\tilde{w}_\infty(\theta; f)$ with bending stiffness, as discussed in Sec. S5.2. Through the angular integral in Eq. (\ref{Seq:converged_PMF_with_angular_weight}), $\tilde{w}_\infty(\theta; f)$ modulates the force contribution to the PMF through its overlap with the combined force--geometry factor $e^{\beta f l \cos{\theta}} \sin{\theta}$. Here, the force-dependent factor $e^{\beta f l \cos{\theta}}$ favors positive-projection orientations with $\theta < 90^\circ$, whereas the geometric factor $\sin{\theta}$ suppresses orientations close to $0^\circ$ and $180^\circ$. In the FJ limit, the angular weight reduces to the constant reference $\tilde{w}_\mathrm{FJ}(\theta; f) = 1$. As illustrated by the representative cases $\beta k_\phi\pi^2 = 10^2$, $1.82 \times 10^3$ and $10^4$ in Fig. \ref{sfig:DBA_angular_weight_vary_kb}, the bulk profile evolves from a single-peaked form centered below $90^\circ$ into two increasingly localized zigzag-compatible branches. As the bending stiffness increases, the emergence of increasingly localization of these branches, together with the displacement of the larger-$\theta$ branch toward and eventually beyond $90^\circ$, alter the overlap with $e^{\beta f l \cos{\theta}} \sin{\theta}$. The resulting force-induced reduction of the PMF is smaller in the barrier region relative to the bonded minimum, thereby increasing $\Delta \mathcal{W}_\infty(f)$. At sufficiently high stiffness, the angular-weight profile approaches a stiffness-independent fixed-bond-angle form, accounting for the corresponding plateau in $\Delta \mathcal{W}_\infty(f)$.

Figs. \ref{sfig:DBA_barrier_measures_vary_kb}b and \ref{sfig:DBA_barrier_measures_vary_kb}c characterize the amplitude and spatial extent of the end-induced barrier perturbation through the terminal barrier reduction $-\delta \mathcal{W}_1(f)$ and the boundary-layer length $\ell_b(f)$, respectively. In the same low-stiffness regime, both measures approach their FJ-limit values. In the FJ limit, the angular weights and activation barriers are independent of bond position, so the terminal barrier coincides with the interior plateau and no terminal boundary layer is present, giving $-\delta \mathcal{W}_1(f) \rightarrow 0$ and $\ell_b(f) \rightarrow 0$. For finite bending stiffness, the truncation of bending correlations at the chain ends makes the angular weights bond-position dependent and generates the near-end structure described in Sec. S5.1. As the bending stiffness increases, the contrast between the near-end and bulk-like angular-weight profiles becomes stronger and remains appreciable over more bond positions (Fig. \ref{sfig:DBA_angular_weight_vary_kb}). Consequently, the terminal barrier becomes progressively lower relative to the interior plateau, while the boundary layer extends farther into the chain, so both $-\delta \mathcal{W}_1(f)$ and $\ell_b(f)$ increase. In the same high-stiffness regime, the near-end angular-weight profiles approach fixed-bond-angle limiting forms, and the two measures correspondingly plateau at $-\delta \mathcal{W}_1(f) \approx 1.59 \times 10^{-2} D_e$ and $\ell_b(f) \approx 31$, respectively.

Overall, increasing bending stiffness raises the bulk-like barrier level while enhancing both the amplitude and spatial extent of the end-induced barrier perturbation, with all three measures evolving from their FJ values toward fixed-bond-angle limits.

\subsection{S6.3 Dependence on applied force}
We then examine how the applied force modifies the long-chain barrier profile, while keeping all remaining parameters fixed at the values used in Sec. S6.1. 

\begin{figure} 
    \includegraphics{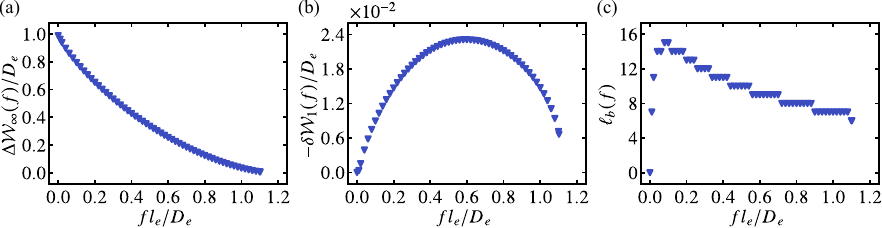}
    \caption{Extracted measures of the long-chain barrier profile as functions of applied force. (a) Interior plateau barrier $\Delta \mathcal{W}_\infty(f)$. (b) Terminal barrier reduction $-\delta \mathcal{W}_1(f)$. (c) Boundary-layer length $\ell_b(f)$. Parameters: $a l_e = 2.15$, $\beta D_e = 279$, $\beta k_\phi\pi^2 = 1.82 \times 10^3$, and $\phi_e = 69^\circ$.}
    \label{sfig:DBA_barrier_measures_vary_f}
\end{figure}

Fig. \ref{sfig:DBA_barrier_measures_vary_f}a shows the interior plateau barrier $\Delta \mathcal{W}_\infty(f)$ as a function of $f l_e/D_e$. At zero force, global rotational symmetry gives the constant angular-weight reference $\tilde{w}_i(\theta; 0) = 1$ for all bond positions. Eq. (\ref{Seq:converged_PMF_with_angular_weight}) then reduces to $\mathcal{W}_i(l;0) = v_\mathrm{str}(l)-\beta^{-1}\ln (2 l^2)$. The radial-entropic contribution $-\beta^{-1}\ln (2 l^2)$ lowers the zero-force activation barrier slightly below the bare dissociation energy $D_e$. At finite force, both the explicit factor $e^{\beta f l \cos{\theta}}$ and the bulk-like angular weight $\tilde{w}_\infty(\theta; f)$ vary with $f$ in the angular integral of Eq. (\ref{Seq:converged_PMF_with_angular_weight}). As discussed in Sec. S5.3, the bulk-like angular-weight profile shows only slight distortions from the flat zero-force reference at $f l_e/D_e = 0.02$, before developing two clearly resolved zigzag-compatible branches at $f l_e/D_e = 0.2$. With increasing force, these branches sharpen and move toward one another. At $f l_e/D_e = 1.0$, the profile is concentrated mainly within the intermediate-orientation range $30^\circ \lesssim \theta \lesssim 110^\circ$, with the outer angular regions strongly depleted. This reshaping changes the overlap of $\tilde{w}_\infty(\theta; f)$ with the combined force--geometry factor $e^{\beta f l \cos{\theta}} \sin{\theta}$, thereby modifying the force contribution to the PMF. Nevertheless, the overall force dependence of the barrier is dominated by the explicit factor $e^{\beta f l \cos{\theta}}$. Because the barrier top occurs at a larger bond length than the bonded minimum, increasing $f$ lowers the PMF more strongly at the barrier than at the minimum. Consequently, $\Delta \mathcal{W}_\infty(f)$ decreases monotonically with increasing force. The calculations extend up to the chain-level critical force $f_c$, defined in Sec. S1.4 as the force at which the metastable minimum and barrier top first coincide for any bond in the chain. For the present parameter set, the terminal bond is limiting because it has the lowest barrier, and the condition $\Delta \mathcal{W}_1(f_c) = 0$ gives $f_c l_e/D_e \approx 1.1$. At this force, the interior plateau barrier remains finite, with $\Delta \mathcal{W}_\infty(f_c) \approx 0.67 \times 10^{-2} D_e$.

Fig. \ref{sfig:DBA_barrier_measures_vary_f}b shows the terminal barrier reduction $-\delta \mathcal{W}_1(f)$ as a function of $f l_e/D_e$. At zero force, the angular weights and activation barriers are independent of bond position, so $-\delta \mathcal{W}_1(0) = 0$. For $0 < f l_e/D_e \lesssim 0.6$, the applied force breaks global rotational symmetry and biases the surrounding subchain conformations. Bending correlations transmit this orientational bias, while their truncation at the chain ends makes the resulting angular-weight profiles bond-position dependent. As shown in Fig. \ref{sfig:DBA_angular_weight_vary_f}, the contrast between the terminal and bulk-like profiles becomes increasingly apparent over this force range. Through the angular integral in Eq. (\ref{Seq:converged_PMF_with_angular_weight}), the two profiles have different overlaps with the combined force--geometry factor $e^{\beta f l \cos{\theta}} \sin{\theta}$, resulting in unequal force-induced reductions of the corresponding PMFs. The terminal barrier therefore decreases more rapidly than the interior plateau, causing $-\delta \mathcal{W}_1(f)$ to increase and reach a maximum of approximately $2.31 \times 10^{-2} D_e$ near $f l_e/D_e \approx 0.6$. Beyond this maximum, the explicit force factor $e^{\beta f l \cos{\theta}}$ increasingly dominates the force dependence of both barriers and produces a rapid overall collapse of the barrier scale. Their absolute difference therefore decreases as $f$ approaches $f_c$, even though the terminal and bulk-like angular-weight profiles remain distinct over this high-force range. At $f_c$, the terminal barrier vanishes while the interior plateau remains finite, giving $-\delta \mathcal{W}_1(f_c) = \Delta \mathcal{W}_\infty(f_c) \approx 0.67 \times 10^{-2} D_e$.

Fig. \ref{sfig:DBA_barrier_measures_vary_f}c shows the boundary-layer length $\ell_b(f)$ as a function of $f l_e/D_e$. At zero force, the position-independent barrier profile contains no terminal boundary layer, so $\ell_b(0) = 0$. Upon applying a weak force, global rotational symmetry is broken and the truncation of bending correlations at the chain ends becomes manifest in the bond-resolved barrier profile. The resulting end-induced deviations remain above the prescribed tolerance over progressively more bond positions, causing $\ell_b(f)$ to rise rapidly and reach a maximum of $15$ bonds near $f l_e/D_e \approx 0.1$. As the force increases further, the explicit force factor $e^{\beta f l \cos{\theta}}$ increasingly dominates the force dependence of the bond-resolved barriers. In particular, the barrier deviations of bonds farther from the chain end decrease in absolute magnitude, shortening the distance over which these deviations remain above the prescribed tolerance. Consequently, $\ell_b(f)$ decreases to $6$ bonds near $f_c$. Because $\ell_b(f)$ is defined on discrete bond positions, this decrease appears as a staircase pattern.

Overall, increasing force monotonically lowers the interior plateau barrier, while the end-induced perturbation varies nonmonotonically in both amplitude and spatial extent. The boundary layer is widest at weak force, reaching $\ell_b = 15$ near $f l_e/D_e \approx 0.1$, whereas the terminal barrier reduction is largest at intermediate force, reaching $-\delta \mathcal{W}_1(f) \approx 2.31 \times 10^{-2} D_e$ neat $f l_e/D_e \approx 0.6$. Thus, the end effect becomes spatially localized before its local amplitude begins to decline as the force approaches $f_c$.

\section{S7 Chain-scission rate}
\subsection{S7.1 Thermodynamic-limit per-bond scission rate}
\begin{figure} 
    \includegraphics{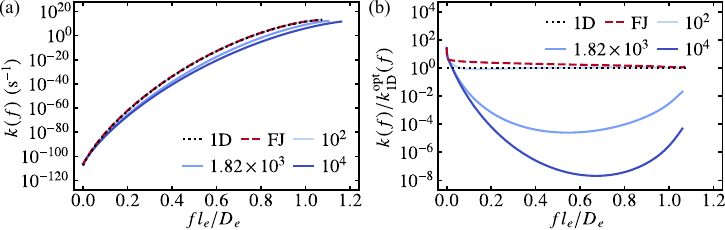}
    \caption{Thermodynamic-limit per-bond scission rate over the full force range. (a) Absolute rates versus dimensionless force $fl_e/D_e$: $k_\mathrm{1D}^\mathrm{opt}(f)$ for the 1D reference, $k_\mathrm{FJ}^\mathrm{opt}(f)$ for the FJ limit, and $k_\infty^\mathrm{opt}(f)$ for finite bending stiffnesses $\beta k_\phi\pi^2 = 10^2$, $1.82 \times 10^3$, and $10^4$. Each curve terminates at its corresponding chain-level critical force $f_c$. (b) The same rates normalized by $k_\mathrm{1D}^\mathrm{opt}(f)$ versus $fl_e/D_e$. Each normalized rate is shown only while both rates entering the ratio remain barrier-defined, terminating at $f_\mathrm{cut} = \min\{f_c, f_c^\mathrm{1D}\}$. Common parameters are $al_e=2.15$ and $\beta D_e=279$; the finite-bending cases use $\phi_e = 69^\circ$, with the kinetic prefactors specified in Sec. S4.}
    \label{sfig:DBA_kbulk_varyf_full}
\end{figure}

As established by the finite-size scaling in the main text, the thermodynamic-limit per-bond scission rate is given by the interior bond rate $k_\infty^\mathrm{opt}(f)$ for finite bending stiffness. For the 1D reference and FJ limit, all bonds $i \geq 2$ share the common rates $k_\mathrm{1D}^\mathrm{opt}(f)$ and $k_\mathrm{FJ}^\mathrm{opt}(f)$, respectively, which therefore define their thermodynamic-limit rates. Accordingly, we denote these rates generically by $k(f)$. Fig. \ref{sfig:DBA_kbulk_varyf_full} compares $k(f)$ for the 1D reference, FJ limit, and finite bending stiffnesses $\beta k_\phi\pi^2 = 10^2$, $1.82 \times 10^3$, and $10^4$.

Fig. \ref{sfig:DBA_kbulk_varyf_full}a shows the absolute rates versus the dimensionless force $fl_e/D_e$. The rates are evaluated from the appropriate full TST expressions for finite bending stiffness, the FJ limit, and the 1D reference (Eqs. (\ref{Sqe:k_bond_with_W}), (\ref{Seq:k_bond_with_W_FJ}), and (\ref{seq:TST_rate_full_1D}), respectively), except for $k_\mathrm{1D}^\mathrm{opt}(0)$, which is defined by Eq. (\ref{seq:zero_force_rate_1D}). We first use the 1D reference to establish the absolute rate scale. At zero force, $k_\mathrm{1D}^\mathrm{opt}(0) \approx 2.30 \times 10^{-108} ~\mathrm{s}^{-1}$. At $f l_e/D_e = 0.6$, it reaches $1.10 \times 10^{-13} ~\mathrm{s}^{-1}$, corresponding to a mean single-bond lifetime of $2.88 \times 10^5~\mathrm{yr}$. At $f l_e/D_e = 0.7$, it further increases to $2.20 \times 10^{-5}~\mathrm{s}^{-1}$, reducing the mean lifetime to $12.63~\mathrm{h}$. As $f$ approaches the 1D critical force $f_c^\mathrm{1D} = a D_e/2$, the rate rises to approximately $1.18 \times 10^{13} ~\mathrm{s}^{-1}$, spanning more than $120$ orders of magnitude from zero force. We therefore restrict the force range shown in Fig. 4 of the main text to $f l_e/D_e \gtrsim 0.6$, since rupture at lower forces occurs on prohibitively long timescales.

The 3D rates exhibit the same rapid force-induced increase as the 1D reference. For $f l_e/D_e \lesssim 0.1$, all curves remain closely grouped on the absolute-rate scale. The FJ and weak-bending ($\beta k_\phi\pi^2 = 10^2$) curves remain close to the 1D curve as the force increases. For $\beta k_\phi\pi^2 = 1.82 \times 10^3$, $k(f)$ becomes substantially suppressed, with the separation from the 1D curve first increasing and then decreasing at higher forces. Increasing the bending stiffness further to $\beta k_\phi\pi^2 = 10^4$ enhances this suppression and widens the separation from the 1D curve at comparable forces. The curves terminate at their respective chain-level critical forces, with $f_c l_e/D_e = 1.075$ for the 1D reference, $1.072$ for the FJ limit, and $1.073$, $1.103$, and $1.161$ for $\beta k_\phi\pi^2 = 10^2$, $1.82 \times 10^3$, and $10^4$, respectively.

Because the quantitative differences are difficult to resolve on the broad absolute-rate scale, Fig. \ref{sfig:DBA_kbulk_varyf_full}b shows the normalized rate $k(f) / k_\mathrm{1D}^\mathrm{opt}(f)$, which removes the common force-acceleration baseline represented by the 1D reference and highlights the conformational effects. At zero force, all 3D cases coincide at $k(0)/k_\mathrm{1D}^\mathrm{opt}(0) \approx 26.7$. This common value follows from global rotational symmetry, which makes the zero-force PMF independent of bending stiffness, while the radial-entropic contribution lowers the 3D activation barrier relative to the 1D reference. n the FJ limit, the normalized rate decreases monotonically toward unity as increasing force confines orientational sampling and makes the FJ response progressively approach the collinear 1D limit. In contrast, all three finite-bending cases show a nonmonotonic force dependence, decreasing to a minimum before turning upward at higher force. For $\beta k_\phi\pi^2 = 10^2$, the minimum is shallow, reaching approximately $0.8$ near $fl_e/D_e \approx 0.14$. With increasing bending stiffness, the minimum becomes much deeper and shifts to higher force, reaching approximately $2.49 \times 10^{-5}$ near $fl_e/D_e \approx 0.54$ for $\beta k_\phi\pi^2 = 1.82\times10^3$ and $2.10 \times 10^{-8}$ near $fl_e/D_e \approx 0.68$ for $\beta k_\phi\pi^2 = 10^4$. This stiffness dependence reflects the evolution of the bulk-like angular weight from a single-peaked profile to increasingly localized and clearly resolved zigzag-compatible branches (Sec. S5.2). The resulting orientational constraints raise the interior plateau barrier with increasing stiffness (Sec. S6.2), thereby progressively lowering the normalized rate. 

The nonmonotonic force dependence can likewise be understood from the force evolution of the angular weight. As the force increases from zero, the bulk-like angular weight evolves from the flat reference to clearly resolved zigzag-compatible branches and then becomes progressively sharper, with the branches moving toward one another at higher force (Sec. S5.3). The emergence of these bending-mediated orientational constraints accounts for the initial decrease of the normalized rate. At higher force, the direct tensile contribution increasingly governs the overall barrier reduction (Sec. S6.3); correspondingly, the relative rate suppression weakens and the normalized rate turns upward.

\subsection{S7.2 Quadratic approximation the force dependence}
As in Sec. S7.1, we denote the thermodynamic-limit per-bond scission rate generically by $k(f)$. To obtain a simple local description of its force dependence, we expand $\ln{k(f)}$ about a reference force $f_\star$,
\begin{equation}
    \ln{k(f)} = \ln{k(f_\star)} + \left.\frac{d \ln{k(f)}}{d f}\right\vert_{f=f_\star} (f - f_\star) + \frac{1}{2}\left.\frac{d^2 \ln{k(f)}}{d f^2}\right\vert_{f=f_\star} (f - f_\star)^2 + \mathcal{O}((f - f_\star)^3).
\end{equation}
Defining
\begin{equation}
    \lambda_\star = \beta^{-1}\left.\frac{d \ln{k(f)}}{d f}\right\vert_{f=f_\star}, \quad
    \kappa_\star = - \beta^{-1}\left.\frac{d^2 \ln{k(f)}}{d f^2}\right\vert_{f=f_\star},
\end{equation}
the quadratic approximation becomes
\begin{equation}
    \label{seq:simple_force_dependence}
    k^{(2)}(f) = k(f_\star) e^{\beta \lambda_\star (f - f_\star) - \frac{1}{2}\beta \kappa_\star (f - f_\star)^2}.
\end{equation}
Here, $\lambda_\star$ is the effective activation length at $f_\star$. Introducing its force-dependent counterpart,
\begin{equation}
    \lambda(f) = \beta^{-1} \frac{d \ln{k(f)}}{d f},
\end{equation}
gives
\begin{equation}
    \kappa_\star = - \left.\frac{d \lambda(f)}{d f}\right\vert_{f=f_\star},
\end{equation}
so $\kappa_\star$ measures the force sensitivity of the effective activation length. In particular, $\kappa_\star > 0$ corresponds to an effective activation length that decreases with increasing force. Eq. (\ref{seq:simple_force_dependence}) provides a local approximation to the full TST rate $k(f)$, with the only approximation arising from truncation of the force expansion at second order.

The quadratic form parallels the extended Bell model \cite{konda2011chemical,makarov2016perspective},
\begin{equation}
    k_\mathrm{EBM}(f) = k_\mathrm{EBM}(0) e^{\beta f \Delta x^\ddagger(0) + \frac{1}{2} \beta f^2 \Delta \chi^\ddagger},
\end{equation}
where $\Delta x^\ddagger(0)$ is the zero-force activation length and $\Delta \chi^\ddagger = \chi_\mathrm{TS} - \chi_\mathrm{R}$ is the transition-state--reactant compliance difference, assumed to be force independent. For the extended Bell model, our coefficients correspond to
\begin{equation}
    \lambda_\star = \Delta x^\ddagger(0) + f_\star \Delta \chi^\ddagger, \quad 
    \kappa_\star = -\Delta \chi^\ddagger
\end{equation}
Our coefficients, however, are obtained directly from the full TST rate at $f_\star$ without assuming a force-independent compliance difference.

To connect these coefficients to the PMF barrier, we consider the high-barrier regime, where the Arrhenius form in Eq. (\ref{Seq:Arrhenius_form}) applies. Neglecting the weaker force dependence of the prefactor then gives
\begin{equation}
    \beta^{-1}\frac{d \ln{k(f)}}{d f} \approx - \frac{d \Delta \mathcal{W}(f)}{df},
\end{equation}
and hence
\begin{equation}
    \lambda_\star \approx - \left.\frac{d \Delta \mathcal{W}(f)}{df}\right\vert_{f=f_\star},\quad
    \kappa_\star \approx \left.\frac{d^2 \Delta \mathcal{W}(f)}{df^2}\right\vert_{f=f_\star}.
\end{equation}
For the 1D reference, $\mathcal{W}_\mathrm{1D}(l; f) = v_\mathrm{str}(l) - f l$, so $\partial \mathcal{W}_\mathrm{1D}/{\partial f} = -l$. Its activation barrier is $\Delta \mathcal{W}_\mathrm{1D}(f) = \mathcal{W}_\mathrm{1D}(l_\mathrm{1D}^\mathrm{b}(f); f) - \mathcal{W}_\mathrm{1D}(l_\mathrm{1D}^\mathrm{m}(f); f)$. Since both locations are stationary points of the PMF, only the explicit force dependence contributes when differentiating the barrier. Hence,
\begin{equation}
    - \frac{d \Delta \mathcal{W}_\mathrm{1D}(f)}{df} = l_\mathrm{1D}^\mathrm{b}(f) - l_\mathrm{1D}^\mathrm{m}(f).
\end{equation}
Evaluating at $f = f_\star$ therefore gives
\begin{equation}
    \lambda_\star \approx l_\mathrm{1D}^\mathrm{b}(f_\star) - l_\mathrm{1D}^\mathrm{m}(f_\star).
\end{equation}
Thus, in the 1D reference, $\lambda_\star$ recovers the conventional activation-length interpretation as the separation between the bonded minimum and barrier top along the bond-length coordinate \cite{makarov2016perspective}. Correspondingly,
\begin{equation}
    \kappa_\star \approx - \frac{d}{df} \left.[l_\mathrm{1D}^\mathrm{b}(f) - l_\mathrm{1D}^\mathrm{m}(f)]\right\vert_{f = f_\star},
\end{equation}
so positive $\kappa_\star$ corresponds to a decreasing separation between the bonded minimum and barrier top with increasing force. In 3D, orientational contributions also affect the force dependence of the PMF barrier, so neither coefficient is determined solely by this bond-length separation.

We choose $f_\star l_e/D_e = 0.7$ and quantify the accuracy of the second-order representation over $0.6 \leq f l_e/D_e \leq 0.8$ by the maximum relative error
\begin{equation}
    \epsilon_\mathrm{max} = 100\% \max_{0.6 \leq f l_e/D_e \leq 0.8} \left|\frac{k^{(2)}(f)}{k(f)} - 1\right|.
\end{equation}
The resulting parameters and errors are summarized in Table \ref{tab:k_expansion_parameters}. Because the representation is local in force, it can be recentered at an appropriate $f_\star$ for other force ranges within the barrier-defined regime, with the corresponding coefficients reevaluated from the full TST rate.

\begin{table*}
\caption{\label{tab:k_expansion_parameters}Parameters of the second-order force representation evaluated at $f_\star l_e/D_e = 0.7$. $k(f_\star)$ is evaluated directly from the full TST expression at $f_\star$, while $\lambda_\star$ and $\kappa_\star$ are obtained by numerical differentiation of the full TST rates. $\epsilon_\mathrm{max}$ is the maximum relative error defined above.}
\begin{ruledtabular}
\begin{tabular}{ccccc}
     Model & $k(f_\star)~(\mathrm{s}^{-1})$ & $\lambda_\star/l_e$ & $\kappa_\star/(\beta l_e^2)$ & $\epsilon_\mathrm{max}$\\ \hline
     1D reference & $2.20 \times 10^{-5}$ & $0.63$ & $4.06\times10^{-3}$ & $6.8\%$\\
     FJ limit & $3.44 \times 10^{-5}$ & $0.63$ & $4.07\times10^{-3}$ &  $14.6\%$\\
     $\beta k_\phi\pi^2 = 10^2$ & $2.37 \times 10^{-5}$ & $0.63$ & $4.08\times10^{-3}$ & $5.5\%$ \\
     $\beta k_\phi\pi^2 = 1.82 \times 10^3$ & $9.01 \times 10^{-10}$ & $0.65$ & $3.55\times10^{-3}$ & $1.7\%$\\
     $\beta k_\phi\pi^2 = 10^4$& $4.74 \times 10^{-13}$ & $0.64$ & $3.14\times10^{-3}$ & $12.8\%$\\
\end{tabular}
\end{ruledtabular}
\end{table*}

\bibliography{bibfile.bib}